\documentclass[onecolumn,draftcls]{IEEEtran}
\usepackage{amsfonts}
\usepackage{amssymb}
\usepackage{dsfont}
\usepackage[dvips]{graphicx}

\usepackage{subfigure}
\usepackage{amsmath}
\newtheorem{theorem}{Theorem}
\newtheorem{lemma}{Lemma}

\begin{document}
\title{Grassmannian Beamforming for MIMO Amplify-and-Forward Relaying}
\author{Behrouz Khoshnevis, Wei Yu, and Raviraj Adve\thanks{The authors are with the
Department of Electrical and Computer Engineering, University of
Toronto, 10 King's College Road, Toronto, Ontario, Canada M5S 3G4
(email: bkhoshnevis@comm.utoronto.ca; weiyu@comm.utoronto.ca;
rsadve@comm.utoronto.ca).}}
\maketitle
\begin{abstract}
In this paper, we derive the optimal transmitter/receiver
beamforming vectors and relay weighting matrix for the
multiple-input multiple-output amplify-and-forward relay channel.
The analysis is accomplished in two steps. In the first step, the
direct link between the transmitter (Tx) and receiver (Rx) is
ignored and we show that the transmitter and the relay should map
their signals to the strongest right singular vectors of the
Tx-relay and relay-Rx channels. Based on the distributions of these
vectors for independent identically distributed (i.i.d.) Rayleigh
channels, the Grassmannian codebooks are used for quantizing and
sending back the channel information to the transmitter and the
relay. The simulation results show that even a few number of bits
can considerably increase the link reliability in terms of bit error
rate. For the second step, the direct link is considered in the
problem model and we derive the optimization problem that identifies
the optimal Tx beamforming vector. For the i.i.d Rayleigh channels,
we show that the solution to this problem is uniformly distributed
on the unit sphere and we justify the appropriateness of the
Grassmannian codebook (for determining the optimal beamforming
vector), both analytically and by simulation. Finally, a modified
quantizing scheme is presented which introduces a negligible
degradation in the system performance but significantly reduces the
required number of feedback bits.
\end{abstract}

\section{Introduction}
The multiple-input multiple-output (MIMO) technology provides a
wireless system with a large number of degrees of freedom, which can
be used for increasing the capacity and/or reliability of the
wireless links. Relaying techniques, on the other hand, can extend
the communication range and coverage, by supporting the shadowed
users through the relay nodes, and reduce the transmission power
required to reach the users far from the base station. These
benefits make MIMO relaying techniques a powerful candidate for
implementation in the next generation of wireless networks.

Considering a system with a single data stream and perfect channel
knowledge at the receiver, several methods can be used to achieve
the benefits of the MIMO link. Maximum ratio transmission and
receiving (MRT-MRC) [1] is one of the simplest methods which can
achieve full diversity order while providing considerable array
gains compared to space-time codes [2]. This gain is obtained at the
expense of the channel knowledge at the transmitter and therefore,
the receiver needs to send the quantized channel information back to
the transmitter. While a general purpose MMSE quantizer can be used
to describe each channel matrix entry, it requires a large number of
feedback bits and does not preserve the structure of the optimal
beamforming vector [3]. A more efficient approach is to have a
common beamforming-vector codebook with finite cardinality and send
back the label of the best beamforming vector to transmitter. This
codebook is designed offline and is known to the transmitter and the
receiver. For the case of flat Rayleigh fading channel, the codebook
design problem has been shown to be related to the Grassmannian line
packing problem [4, 5, 6].

In this paper, we generalize the idea of MRT-MRC to a MIMO link with
an amplify-and-forward relay station. The scenario, considered in
this paper, comprises a transmitter (Tx), a receiver (Rx) and a
relay which helps the transmitter to send its data to the receiver.
A general information theoretic analysis of MIMO relay link has been
presented in [7] and [8]. Although an efficient signaling through
the relay channel requires a full-duplex relay with specific
processing capabilities (e.g. encoding/decoding),
amplify-and-forward (AF) relays are still attractive due to their
lower complexity. Moreover, the full-duplex assumption cannot be
realized by the current technology, as the input and output signals
need to be separated in time or frequency at the relay. For these
reasons, this paper focuses on the half-duplex AF relay system. In
such a system, the transmitter sends out its symbol in the first
time slot and the relay and the receiver receive their signal. In
the second time slot, the transmitter remains silent and the relay
multiplies its received signal by a matrix (amplification) and sends
the resulting signal to the receiver. The receiver decodes the
transmitted symbol based on the signals received in two consecutive
slots.

The half-duplex MIMO AF scenario has been considered in [9] and
[10], where the authors present different solutions for maximizing
the instantaneous capacity with respect to the weighting
(amplification) matrix of the relay. These papers assume no channel
state information at the transmitter (CSIT) and consider uniform
power allocation over the Tx antennas. The work in [11] considers
the same problem with perfect CSIT and derives the optimal power
allocation scheme for the transmitter and relay (without considering
the Tx-Rx link). Our problem setup is different from these papers in
two major aspects, listed below:
\begin{itemize}
\item The objective of the aforementioned references is the maximization of the
instantaneous capacity. Our problem, however, can be categorized as
a beamforming problem, where we optimize the Tx/Rx beamforming
vectors and the relay matrix to maximize the signal-to-noise ratio
(SNR) of a single data stream at the Rx output.
\item The above papers assume either no channel information or complete
channel information at the transmitter or the relay. Our work,
however, focuses on a ``limited feedback'' system, where the
receiver end of a link sends the properly quantized channel
information back to the transmitter end.
\end{itemize}

The analysis in this paper starts by first ignoring the direct link
between transmitter and receiver, where we show that the transmitter
and the relay should map their symbols to the strongest right
singular vectors of the Tx-relay and relay-Rx channels. For Rayleigh
fading channels, these vectors are uniformly distributed on the unit
sphere and therefore the Grassmannian criterion can be used
separately for Tx-relay and relay-Rx codebook design.

In the second part of the paper, we include the direct link in the
system model. As expected, one needs to know both Tx-relay and Tx-Rx
channel matrices to determine the optimal Tx beamforming vector for
this case. We first assume that such a knowledge is available (for
example at the relay), and we derive the optimization problem that
characterizes the optimal Tx beamforming vector. Although this
problem does not appear to have an analytic solution, we are able to
show that for i.i.d. Rayleigh channels the solution to this problem
is uniformly distributed on the unit sphere, based on which, the
appropriateness of the Grassmannian quantizer can be shown
analytically.

In the next step, we relax the assumption of complete knowledge of
the Tx-relay and Tx-Rx channels. Without this assumption, the Rx and
relay should somehow exchange their information of the Tx-relay and
Tx-Rx channels. We focus on a scheme, where the Rx quantizes the
Tx-Rx channel matrix and sends it to the relay, which already knows
the Tx-relay channel matrix. Assuming an ideal scalar quantizer for
the singular values of the Tx-Rx channel matrix, we justify the use
of the Grassmannian quantizer for quantizing the singular vectors.
Finally, we present a modified quantizer, which only quantizes the
strongest singular vector of the Tx-Rx channel and sends it to the
relay. This quantizer requires fewer number of feedback bits and
performs very close to the original quantizer.

The remainder of this paper is organized as follows. In Section II,
we present a brief introduction to Grassmannian line packing problem
and its connection to the MIMO beamforming codebook design. Section
III presents the problem setup and the solution for the MIMO relay
channel without considering the direct link. In Section IV, the
beamforming codebook design problem is solved with the direct link
included in the system model. The simulation results are discussed
in Section V. Finally, Section VI concludes the paper.

\emph{Notations:}  $\mathds{R}$ and $\mathds{C}$ denote the set of
real and complex numbers. Bold upper case and lower case letters
denote matrices and vectors. $\mathbf{I}$ shows the identity matrix.
$\mathcal{U}^m$ denotes the set of all unitary matrices in
$\mathds{C}^{m\times m}$. $~~|\cdot|$ and $\|\cdot\|$ show the
absolute value of a scalar and the Euclidean norm of a vector.
$\|\cdot\|_{_F}$ denotes the Frobenius norm of a
matrix\footnote{$\|\mathbf{A}\|_{_F}^2=\sum_{i,j}{|a_{ij}|^2}=\textmd{Trace}(\mathbf{A}\mathbf{A}^H)=\sum_{k}{\sigma_k^2}$,
where $\sigma_k$'s are the singular values of the matrix
$\mathbf{A}=[a_{ij}]$.}. $(\cdot)^T$ and $(\cdot)^H$ denote the
transpose and Hermitian of a matrix. The notation
$\mathbf{\Phi}=\texttt{diag}_{m\times
n}(\phi_1,\phi_2,\cdots,\phi_r)$ with $r=\min\{m,n\}$ shows a
rectangular diagonal matrix $\mathbf{\Phi}\in\mathds{C}^{m\times n}$
with $\mathbf{\Phi}(i,i)=\phi_i$ for $i=1,2,\cdots,r$ and
$\mathbf{\Phi}(i,j)=0$ for $i\neq j$. For an arbitrary matrix
$\mathbf{H}\in\mathds{C}^{m\times n}$, the singular value
decomposition (SVD) of $\mathbf{H}$ is expressed as
$\mathbf{H}=\mathbf{U}\mathbf{\Sigma}\mathbf{V}^H$, where
$\mathbf{U}\in\mathcal{U}^m$ and $\mathbf{V}\in\mathcal{U}^n$
include the left and right singular vectors as their columns, and
$\mathbf{\Sigma}=\texttt{diag}_{m\times
n}(\sigma_1,\sigma_2,\cdots,\sigma_r)$, where $r=\min\{m,n\}$ and
$\sigma_1{\geq}\sigma_2{\geq}\cdots{\geq}\sigma_r{\geq} 0$; if
$R=\texttt{rank}(\mathbf{H})$, the first $R$ nonzero diagonal
enteries of $\mathbf{\Sigma}$ are called the singular values of
$\mathbf{H}$. $\mathcal{CN}(0,\mathbf{\Sigma})$ represents a
circularly symmetric complex Gaussian distribution with zero mean
and covariance matrix $\mathbf{\Sigma}$. Finally,
$\mathrm{E}\{\cdot\}$ denotes the expectation operation.

\section{MIMO Beamforming Codebook Design and Grassmannian Line
Packing}

\begin{figure}
\centering
\includegraphics[width=3.0in]{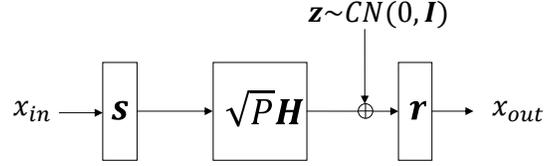}
\caption{Single stream MIMO link with Tx and Rx beamforming.}
\end{figure}

The connection between Grassmannian line packing problem and
beamforming codebook design for a Rayleigh fading channel has been
independently observed in [5] and [6]. Consider the MIMO channel in
Fig. 1. The transmitter maps the symbol $x_{in}$ to the antenna
array using the beamforming vector $\mathbf{s}$. The signal passes
through the channel $\sqrt{P}\mathbf{H}$ with complex Gaussian noise
$\mathbf{z}\sim \mathcal{CN}(0,\mathbf{I})$. The receiver recovers
the symbol $x_{out}$ using the receive beamforming vector
$\mathbf{r}$. The matrix $\sqrt{P}\mathbf{H}\in\mathds{C}^{l\times
m}$ models the flat fading channel and $m$ and $l$ are the number of
the Tx and Rx antennas respectively. The entries of $\mathbf{H}$ are
assumed to be independent and identically distributed according to
$\mathcal{CN}(0,1)$. The coefficient $P$ is referred to as the
``link signal-to-noise ratio (SNR)''. The output symbol can be
expressed as
\[x_{out}=\sqrt{P}\mathbf{r}^H\mathbf{H}\mathbf{s}x_{in}+\mathbf{r}^H\mathbf{z}.\]
Assuming a transmission power constraint of $1$, satisfied by
$\mathrm{E}\{|x_{in}|^2\}=1$ and $\|\mathbf{s}\|=1$, the received
SNR $\gamma$ is:
\[ \gamma=\frac{P|\mathbf{r}^H\mathbf{Hs}|^2}{\|\mathbf{r}\|^2},
\]
which should be maximized with respect to $\mathbf{r}$ and
$\mathbf{s}$. Maximization with respect to $\mathbf{r}$ is achieved
by matching $\mathbf{r}=\mathbf{Hs}$, hence the optimal $\mathbf{s}$
should maximize $\gamma=P\|\mathbf{Hs}\|^2$. It is easy to show that
the optimal $\mathbf{s}$ is the right singular vector of
$\mathbf{H}$ corresponding to its largest singular value. If we
denote the largest singular value and the corresponding right
singular of $\mathbf{H}$ by $\sigma_1$ and $\mathbf{v}_1$,  the
optimal Tx beamforming vector is equal to
$\mathbf{s}^\star=\mathbf{v}_1$ and the maximum SNR is
$\gamma^\star=P\sigma_1^2$.


For the Rayleigh fading channel matrix $\mathbf{H}$, the singular
vectors have been shown to be uniformly distributed on the unit
sphere in $\mathds{C}^{m}$ (see [5], [12]). Therefore, a good
quantizer of the optimal $\mathbf{s}$, in a sense, should place its
codebook vectors uniformly on the unit sphere. This requirement can
be shown to be related to the criterion used in the Grassmannian
line packing problem, which we describe next.

Consider the complex space $\mathds{C}^m$ and let $\Omega$ be the
unit sphere,
$\Omega=\{\mathbf{w}\in\mathds{C}^m|\|\mathbf{w}\|=1\}$. Define the
distance of two unit vectors to be sine of the angle between them:
\begin{equation}
d(\mathbf{w}_1,\mathbf{w}_2)=\sqrt{1-|\mathbf{w}_1^H\mathbf{w}_2|^2},
\end{equation}
for $\mathbf{w}_1,\mathbf{w}_2\in\Omega$. For a codebook
$\mathbf{C}=\{\mathbf{w}_1,\mathbf{w}_2,\cdots ,\mathbf{w}_N\}$ with
$N$ distinct unit vectors, define $\delta(\mathbf{C})$ as the
minimum distance of the codebook:
\[\delta(\mathbf{C})=\min_{\substack{\mathbf{w}_i,\mathbf{w}_j\in \mathbf{C}\\ i\neq j}}{d(\mathbf{w}_i,\mathbf{w}_j)}.\]
For a fixed dimension $m$ and codebook size $N$, the Grassmannian
line packing problem [4] is that of finding a codebook $\mathbf{C}$
of size $N$ with the largest minimum distance. Many researchers have
studied the solution to this problem for moderate values of $m$ and
$N$ [13], [14]. However, there is no known standard way of finding
these codebooks in general.


For the problem setup in Fig. 1, consider a beamforming codebook
$\mathbf{C}(N,\delta)$ of size $N$ and minimum distance $\delta$.
The receiver chooses the vector in this codebook that maximizes the
SNR and sends the label of this vector back to the transmitter. Let
$\tilde{\gamma}$ denote the resulting SNR:
$\tilde{\gamma}=\max_{\mathbf{w}\in\mathbf{C}}{P\|\mathbf{H}\mathbf{w}\|^2}$.
The authors in [5] have used the distribution of optimal beamforming
vector $\mathbf{s}^\star$ to bound the average SNR loss as:
\begin{IEEEeqnarray}{lll}
\mathrm{E}\{\gamma^\star\}&-&\mathrm{E}\{\tilde{\gamma}\}  ~\leq  \nonumber\\
&P&\mathrm{E}\{\sigma_1^2\}\left(
1{-}N\left(\frac{\delta}{2}\right)^{2(m{-}1)}\left(1{-}\frac{\delta^2}{4}\right)\right),
\end{IEEEeqnarray}
where $m$ is the space dimension (number of Tx antennas). The upper
bound in (2) is a decreasing function of $\delta$, for any $m>1$.
Therefore, to minimize the upper bound of the SNR loss, we should
maximize the minimum distance of the codebook. This is the same
criterion used in the definition of the Grassmannian line packing
problem and establishes the connection between the beamforming
codebook design problem and the Grassmannian line packing.

Before concluding this section, we mention that the codebook design
problem for the beamforming system in Fig. 1 has been generalized by
[15] to the multiplexing systems, where the Tx transmits multiple
substreams to the Rx. In such systems, the transmitter and receiver
share a codebook of precoding matrices and the receiver sends back
the label of the matrix that maximizes a certain performance
criterion (e.g. the minimum substream SNR). In this paper, we take
the first step in designing the limited feedback systems for
beamforming over MIMO AF relay channels. The generalization of the
relay problem to the case of multiple data streams is considered as
the future work.

\section{MIMO Amplify and Forward Relay Channel without the Direct Link}

In this section, we consider the MIMO amplify-and-forward (AF) relay
channel without the direct link and derive the optimal
transmitter/receiver beamforming vectors and relay weighting matrix
in Subsection III.A. Next, we present the quantization scheme in
Subsection III.B. It should be noted that if the relay performs
decode-and-forward, the MIMO relay channel reduces to two MIMO links
in series, therefore the optimal structure and the quantization
scheme in Section I can be applied to each of the links separately.
However, the derivation of the optimal unquantized scheme and
designing the corresponding quantization scheme is not trivial when
the relay performs amplify-and-forward.

\subsection{Optimal Unquantized Scheme}
Consider the MIMO amplify-and-forward relay system in Fig. 2a, where
the direct link between transmitter and receiver is ignored. The
transmitter, the relay and the receiver are equipped with $m$, $n$
and $l$ antennas, respectively. The matrices
$\sqrt{P_1}\mathbf{H}_1\in\mathds{C}^{n\times m}$ and
$\sqrt{P_2}\mathbf{H}_2\in\mathds{C}^{l\times n}$ model the flat
fading channels of the Tx-relay and relay-Rx links, respectively.
The coefficients $P_1$ and $P_2$ are referred to as Tx-relay and
relay-Rx ``link SNRs''. The transmitter uses the vector $\mathbf{s}$
for beamforming. The relay multiplies its noisy received signal by
the matrix $\mathbf{W}\in\mathds{C}^{l\times l}$ and sends it to the
receiver. The receiver recovers its symbol using the receive
beamforming (combining) vector $\mathbf{r}$. We assume power
constraints equal to $1$ at the transmitter and the relay outputs.

\begin{figure}%
\centering \subfigure[]{
\includegraphics[width=3.5in]{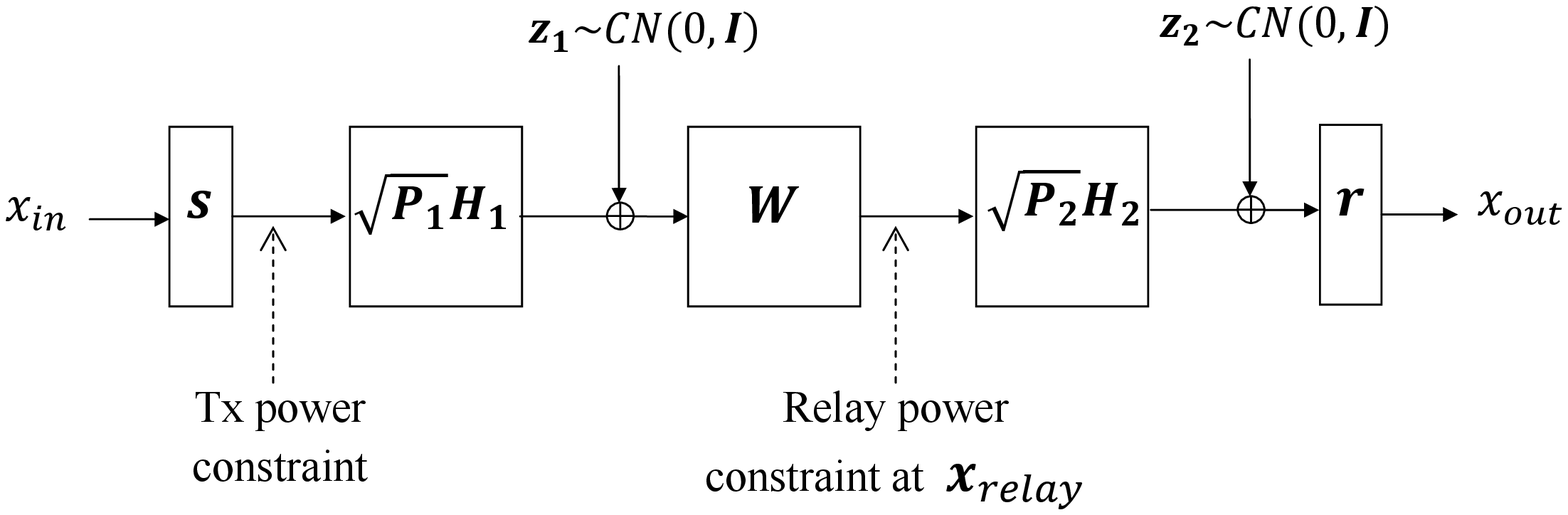}}\\
\subfigure[]{
\includegraphics[width=2.5in]{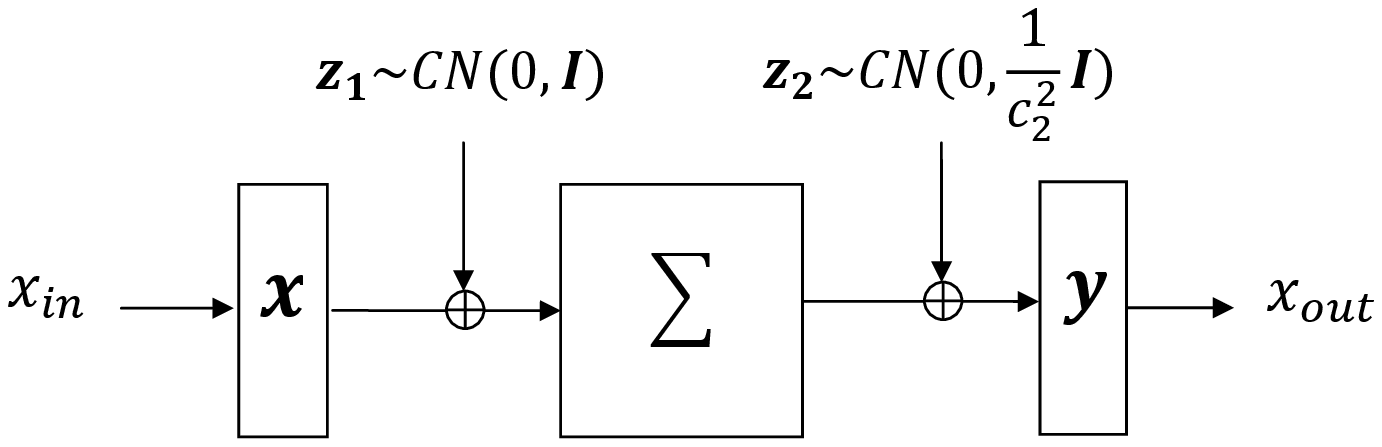}}
\caption[]{(a) MIMO amplify-and-forward relay channel model without
the direct link, (b) The model after the change of variables
$\mathbf{x}=\sqrt{P_1}\mathbf{V}^H\mathbf{H}_1\mathbf{s}$ and
$\mathbf{y}=\sqrt{P_2}\mathbf{U}^H\mathbf{H}_2^H\mathbf{r}$.}
\end{figure}

The problem is to find the optimal $\mathbf{s}$, $\mathbf{W}$ and
$\mathbf{r}$, to maximize the SNR at the receiver output subject to
power constraints at the Tx and at the relay. For this problem
setup, a reasonable solution is ``matching'', as described below.
The transmitter should map its symbol to the strongest right
singular vector of $\mathbf{H}_1$ (as described in Section II). The
relay should absorb maximum signal power by matching to the
effective channel\footnote{This matching vector is parallel to the
strongest left singular vector of $\mathbf{H}_1$.}
$\textbf{H}_1\textbf{s}$, scale the resulting (noisy) signal to meet
its power constraint and transmit it through the strongest right
singular vector of $\mathbf{H}_2$. Finally, the receiver should
match to the relay-Rx link by using the strongest left singular
vector of $\mathbf{H}_2$ as the Rx beamformer. This matching
solution is depicted in Fig. 3a, in which
\begin{IEEEeqnarray}{lll}
\mathbf{H}_1=\mathbf{A}\mathbf{\Phi}\mathbf{B}^H,\nonumber\\
\mathbf{H}_2=\mathbf{F}\mathbf{\Psi}\mathbf{G}^H,
\end{IEEEeqnarray}
are the SVD decompositions of $\mathbf{H}_1$ and $\mathbf{H}_2$, and
\begin{IEEEeqnarray}{lllllll}
\mathbf{A}&{=}&[\mathbf{a}_1|\mathbf{a}_2|\cdots|\mathbf{a}_n]\in\mathcal{U}^n,& &\mathbf{F}&{=}&[\mathbf{f}_1|\mathbf{f}_2|\cdots|\mathbf{f}_l]\in\mathcal{U}^l,\nonumber\\
\mathbf{B}&{=}&[\mathbf{b}_1|\mathbf{b}_2|\cdots|\mathbf{b}_m]\in\mathcal{U}^m,& &\mathbf{G}&{=}&[\mathbf{g}_1|\mathbf{g}_2|\cdots|\mathbf{g}_n]\in\mathcal{U}^n,\nonumber\\
\mathbf{\Phi}&{=}&\texttt{diag}_{n{\times}
m}\{\phi_1,\phi_2,{\cdots},\phi_{r_1}\},&
&\mathbf{\Psi}&{=}&\texttt{diag}_{l{\times}
n}\{\psi_1,\psi_2,{\cdots},\psi_{r_2}\},\nonumber
\end{IEEEeqnarray}
where $r_1=\min\{n,m\}$, $r_2=\min\{l,n\}$. Although matching seems
to be the natural solution to this problem, showing that the optimal
$\mathbf{W}$ is a rank one matrix and that matching is optimal is
not trivial. This is mainly due to the noise amplification at the
relay, which generates colored noise at the receiver input. In the
remainder of this section, we present a proof for the optimality of
this scheme.

The relay and receiver output signals in Fig. 2a are:
\begin{IEEEeqnarray}{lll}
x_{out}&=&\sqrt{P_1P_2}\mathbf{r}^H\mathbf{H}_2\mathbf{W}\mathbf{H}_1\mathbf{s}x_{in}+\sqrt{P_2}\mathbf{r}^H\mathbf{H}_2\mathbf{W}\mathbf{z}_1+\mathbf{r}^H\mathbf{z}_2,\nonumber\\
\mathbf{x}_{relay}&=&\sqrt{P_1}\mathbf{W}\mathbf{H}_1\mathbf{s}x_{in}+\mathbf{W}\mathbf{z}_1,\nonumber
\end{IEEEeqnarray}
where  $\mathbf{z}_1\sim \mathcal{CN}(0,\mathbf{I})$ and
$\mathbf{z}_2\sim \mathcal{CN}(0,\mathbf{I})$ are the complex
Gaussian noise vectors at the relay and Rx input. The transmitter
power constraint is satisfied by letting
$\mathrm{E}\{|x_{in}|^2\}=1$ and $\|\mathbf{s}\|=1$. Also, the relay
power constraint, which limits the power of the amplified signal and
noise, can be expressed as:\[
\mathrm{E}\{\left\|{\mathbf{x}_{relay}}\right\|^2\}=P_1\left\|\mathbf{W}\mathbf{H}_1\mathbf{s}\right\|^2+\|\mathbf{W}\|_{_F}^2=1.\]
Finally, the ``received SNR'' can be written as:
\[
\gamma=\frac{P_1P_2\left|\mathbf{r}^H\mathbf{H}_2\mathbf{W}\mathbf{H}_1\mathbf{s}\right|^2}{P_2\left\|\mathbf{W}^H\mathbf{H}_2^H\mathbf{r}\right\|^2+\|\mathbf{r}\|^2},
\]
where we can assume $\|\mathbf{r}\|=1$, without loss of generality.
The optimization problem can be summarized as:
\begin{IEEEeqnarray}{ll}
\max~~&{\frac{P_1P_2\left|\mathbf{r}^H\mathbf{H}_2\mathbf{W}\mathbf{H}_1\mathbf{s}\right|^2}{P_2\left\|\mathbf{W}^H\mathbf{H}_2^H\mathbf{r}\right\|^2+1}}\\
\textmd{s.t.}&\nonumber\\
&\left\{\begin{array}{ll} \|\mathbf{s}\|=\|\mathbf{r}\|=1\nonumber\\
P_1\left\|\mathbf{W}\mathbf{H}_1\mathbf{s}\right\|^2+\|\mathbf{W}\|_{_F}^2=1
\nonumber\\{\mathbf{W}\in\mathds{C}^{l\times l},~
\mathbf{s}\in\mathds{C}^m,~ \mathbf{r}\in\mathds{C}^n}.\nonumber
\end{array}\right.
\end{IEEEeqnarray}

\begin{theorem}
The optimal values of Tx/Rx beamforming vectors and relay weighting
matrix for the SNR maximization problem in (4) are given by:
\begin{equation}
\mathbf{s}^\star=\mathbf{b}_1,~~\mathbf{r}^\star=\mathbf{f}_1,~~\mathbf{W}^\star=\sigma\mathbf{g}_1\mathbf{a_1}^H,\nonumber
\end{equation}
where we have used the SVD equations in (3), and
$\sigma={\left({1+P_1\phi_1^2}\right)}^{-\frac{1}{2}}$. Note that
the optimal weighting matrix $\mathbf{W}^\star$ is a rank one
matrix.
\end{theorem}

\begin{proof}
The optimization is accomplished in two steps. In the first step, we
fix $\mathbf{s}$ and $\mathbf{r}$ and maximize the objective with
respect to $\mathbf{W}$. In the second step, optimal $\mathbf{s}$
and $\mathbf{r}$ are derived after substituting the optimal
$\mathbf{W}$ in the SNR expression.

\emph{\textbf{Step 1)} Maximization with respect to $\mathbf{W}$:}

Define $\mathbf{h}_1=\sqrt{P_1}\mathbf{H}_1\mathbf{s}$ and
$\mathbf{h}_2=\sqrt{P_2}\mathbf{H}_2^H\mathbf{r}$. By fixing
$\mathbf{s}$ and $\mathbf{r}$, $\mathbf{h}_1$ and $\mathbf{h}_2$ are
also fixed. Let $c_1=\left\|\mathbf{h}_1\right\|$ and
$c_2=\left\|\mathbf{h}_2\right\|$.

Consider $\mathbf{W}=\mathbf{U}\mathbf{\Sigma}\mathbf{V}^H$ as the
SVD of $\mathbf{W}$, where $\mathbf{U},\mathbf{V}\in\mathcal{U}^l$
and $\mathbf{\Sigma}=\texttt{diag}_{l\times
l}\{\sigma_1,\sigma_2,\cdots,\sigma_l\}$. The calculations provided
below perform the optimization with respect to $\mathbf{U}$,
$\mathbf{V}$ and $\mathbf{\Sigma}$.

Define $\mathbf{x}=\mathbf{V}^H\mathbf{h}_1$ and
$\mathbf{y}=\mathbf{U}^H\mathbf{h}_2$, which impose the constraints
$\|\mathbf{x}\|=\left\|\mathbf{h}_1\right\|=c_1$ and
$\|\mathbf{y}\|=\left\|\mathbf{h}_2\right\|=c_2$ on
$\mathbf{x}=\left[x_1,x_2,\cdots,x_l\right]^T,\mathbf{y}=\left[y_1,y_2,\cdots,y_l\right]^T\in
\mathds{C}^l$. The maximization with respect to $\mathbf{U}$,
$\mathbf{V}$ and $\mathbf{\Sigma}$, i.e. (4), can now be rephrased
as a maximization with respect to $\mathbf{x}$, $\mathbf{y}$ and
$\mathbf{\Sigma}$:
\begin{IEEEeqnarray}{lll}
&\max~~&{\frac{\left|\mathbf{y}^H\mathbf{\Sigma}\mathbf{x}\right|^2}{\left\|\mathbf{\Sigma}\mathbf{y}\right\|^2+1}}\\
&\textmd{s.t.}&\nonumber\\
&&\left\{\begin{array}{ll} \|\mathbf{x}\|=c_1\nonumber\\\|\mathbf{y}\|=c_2\nonumber\\
\sum_{i=1}^l{\sigma_i^2\left|x_i\right|^2}+\sum_{i=1}^l{\sigma_i^2}=1\nonumber\\x_i,~y_i\in\mathds{C},~\sigma_i\geq0,~i=1,2,\cdots,l\nonumber
\end{array}\right.
\end{IEEEeqnarray}
where the power constraint of the relay is computed as
follows:\[\begin{array}{lll}
P_1\left\|\mathbf{W}\mathbf{H}_1\mathbf{s}\right\|^2+\|\mathbf{W}\|_{_F}^2&=&\left\|\mathbf{U}\mathbf{\Sigma}\mathbf{V}^H\mathbf{h}_1\right\|^2+\|\mathbf{W}\|_{_F}^2\nonumber\\
&=&\left\|\mathbf{\Sigma}\mathbf{x}\right\|^2+\sum_{i}{\sigma_i^2}\nonumber\\&=&\sum_{i}{\sigma_i^2\left|x_i\right|^2}+\sum_{i}{\sigma_i^2}.\nonumber\end{array}\]
The problem in (5) is exactly the SNR maximization problem for the
(single-hop) MIMO link depicted in Fig. 2b, where $\mathbf{x}$ and
$\mathbf{y}$ are the transmit and receive beamformers and
$\mathbf{\Sigma}$ is the channel. Note that the only constraint on
the receiver beamformer $\mathbf{y}$ is on its Euclidean norm,
therefore, the optimal $\mathbf{y}$ is the minimum mean square error
(MMSE) filter\footnote{For a general input-output relation
$x_{out}=\mathbf{y}^H(\mathbf{h}x_{in}+\mathbf{z})$, the optimal
(SNR maximizing) receiver beamforming vector is the MMSE filter
$\mathbf{y}=c\mathbf{K}^{-1}\mathbf{h}$ for $\mathbf{K}$ being the
covariance matrix of $\mathbf{z}$ and any scalar $c$. The resulting
(maximum) SNR is $\gamma=\mathbf{h}^H\mathbf{K}^{-1}\mathbf{h}$.}.
Hence, the optimal $\mathbf{y}$ and the corresponding SNR are:
\begin{IEEEeqnarray}{lll}
\mathbf{y}&=&c\left(\mathbf{\Sigma}^2+\frac{1}{c_2^2}\mathbf{I}\right)^{-1}\mathbf{\Sigma
x}\\
\gamma&=&\mathbf{x}^H\mathbf{\Sigma}\left(\mathbf{\Sigma}^2+\frac{1}{c_2^2}\mathbf{I}\right)^{-1}\mathbf{\Sigma
x},
\end{IEEEeqnarray}
where $\mathbf{\Sigma}^2+\frac{1}{c_2^2}\mathbf{I}$ is the
covariance matrix of the equivalent noise and $\mathbf{\Sigma x}$ is
the equivalent channel from the input symbol to the receiver input.
The scalar $c$ is chosen to satisfy the constraint
$\|\mathbf{y}\|=c_2$.

For the next step, we find an upper bound for the SNR expression in
(7) by considering the constraints on $x_i$'s and $\sigma_i$'s, and
we present the optimal values of $\mathbf{x}$ and $\mathbf{\Sigma}$
that achieve this upper bound. Considering (7), we get to the
following maximization problem.
\begin{IEEEeqnarray}{ll}
\max~~&\sum_{i=1}^{l}{|x_i|^2\frac{\sigma_i^2}{\sigma_i^2+\frac{1}{c_2^{2}}}}\\
\textmd{s.t.}&\nonumber\\
&\left\{
\begin{array}{ll}
\|\mathbf{x}\|=c_1\nonumber\\
\sum_{i=1}^{l}{\sigma_i^2\left|x_i\right|^2}+\sum_{i=1}^l{\sigma_i^2}=1\\
x_i\in\mathds{C},~\sigma_i\geq0,~i=1,2,\cdots,l\nonumber
\end{array}
\right.
\end{IEEEeqnarray}

Define
$\beta_i=\frac{|x_i|^2}{c_1^2}=\frac{|x_i|^2}{\|\mathbf{x}\|^2}$.
Clearly, $0\leq\beta_i\leq 1$ and $\sum_{i=1}^{l}{\beta_i}=1$. Now,
consider the objective function in (8):
\begin{IEEEeqnarray}{lcl}
\gamma&=&\sum_{i}{|x_i|^2\frac{\sigma_i^2}{\sigma_i^2+{1/{c_2^2}}}}=c_1^2\sum_{i}{\frac{|x_i|^2}{c_1^2}\frac{\sigma_i^2}{\sigma_i^2+{1/{c_2^2}}}}\nonumber\\
&=&c_1^2\sum_{i}{\beta_i\frac{\sigma_i^2}{\sigma_i^2+{{1}/{c_2^{2}}}}}\leq
c_1^2\frac{\sum_{i}{\beta_i\sigma_i^2}}{\sum_{i}{\beta_i\sigma_i^2}+{{1}/{c_2^2}}}\\
&=&c_1^2\frac{\sum_{i}{\sigma_i^2|x_i|^2}}{\sum_{i}{\sigma_i^2|x_i|^2}+{{c_1^2}/{c_2^2}}}=c_1^2\frac{\zeta}{\zeta+{{c_1^2}/{c_2^2}}},
\end{IEEEeqnarray}
where $\zeta\stackrel{def}{=}\sum_{i}{\sigma_i^2|x_i|^2}$. The
inequality in (9) is a result of the concavity of the function
$\frac{t}{t+{1/c_2^2}}$ for $t\geq 0$.

Now, from the second constraint of the problem (8), we have:
\[1{-}\sum_{i}{\sigma_i^2}=\sum_{i}{\sigma_i^2|x_i|^2}\leq\sum_{i}{\sigma_i^2}~\cdot~\sum_{i}{|x_i|^2}
=c_1^2\sum_{i}{\sigma_i^2}.\] Therefore,
$\sum_{i}{\sigma_i^2}\geq\frac{1}{1+c_1^2}$ and by applying the same
constraint, we can bound $\zeta$:
\begin{equation}
\zeta=\sum_{i}{\sigma_i^2|x_i|^2}=1-\sum_{i}{\sigma_i^2}\leq\frac{c_1^2}{1+c_1^2}.
\end{equation}

Finally, by combining (10) and (11), and noting that (10) is
increasing in $\zeta$, we have the following upper bound for the
SNR:
\begin{equation}
\gamma\leq\frac{c_1^2c_2^2}{1+c_1^2+c_2^2}.
\end{equation}

By reconsidering the problem in (5), it is easy to check that the
following choices of $\mathbf{x}$, $\mathbf{\Sigma}$ and
$\mathbf{y}$ satisfy the constraints and achieve the upper bound in
(12).
\begin{equation}
\mathbf{x}{=}{[}c_1{,}0{,}{\cdots}{,}0{]}^T,~~\mathbf{y}{=}{[}c_2{,}0{,}{\cdots}{,}0{]}^T,~~\mathbf{\Sigma}{=}\texttt{diag}_{l\times
l}{\{}\sigma{,}0{,}{\cdots},{0}{\}},
\end{equation}
where $\sigma=\left(1+c_1^2\right)^{-\frac{1}{2}}$. Recalling the
definitions of $\mathbf{x}$, $\mathbf{y}$, $c_1$ and $c_2$, the
optimal values in (13) can be achieved by:
\begin{IEEEeqnarray}{lll}
\mathbf{V}{=}{[}\hat{\mathbf{h}}_1{|}\mathbf{v}_1{|}\cdots|\mathbf{v}_{l{-}1}{]},~~\mathbf{U}{=}{[}\hat{\mathbf{h}}_2{|}\mathbf{u}_1{|}\cdots{|}\mathbf{u}_{l{-}1}{]},\nonumber\\
\mathbf{\Sigma}{=}\texttt{diag}_{l\times
l}{\{}\sigma{,}0{,}{\cdots},{0}{\}}
\end{IEEEeqnarray}
where
$\hat{\mathbf{h}}_1=\frac{\mathbf{h}_1}{\left\|\mathbf{h}_1\right\|}$,
$\hat{\mathbf{h}}_2=\frac{\mathbf{h}_2}{\left\|\mathbf{h}_2\right\|}$
and $\sigma=(1+\left\|\mathbf{h}_1\right\|^2)^{-\frac{1}{2}}$. Here
$\{\mathbf{v}_1,\cdots,\mathbf{v}_{l-1}\}$ and
$\{\mathbf{u}_1,\cdots,\mathbf{u}_{l-1}\}$ are arbitrary orthonormal
basis for the null-spaces of the $\mathbf{h}_1$ and $\mathbf{h}_2$
respectively.

To summarize, having $\mathbf{s}$ and $\mathbf{r}$ fixed, the
optimal structure of
$\mathbf{W}=\mathbf{U}\mathbf{\Sigma}\mathbf{V}^H$ and the
corresponding SNR value are:
\begin{IEEEeqnarray}{l}
\mathbf{W}=\sigma\hat{\mathbf{h}}_2{\hat{\mathbf{h}}_1}^H\\
\gamma=\frac{{\left\|\mathbf{h}_1\right\|}^2{\left\|\mathbf{h}_2\right\|}^2}{1+{\left\|\mathbf{h}_1\right\|}^2+{\left\|\mathbf{h}_2\right\|}^2},
\end{IEEEeqnarray}
where $\sigma=(1+\left\|\mathbf{h}_1\right\|^2)^{-\frac{1}{2}}$,
$\mathbf{h}_1=\sqrt{P_1}\mathbf{H}_1\mathbf{s}$, and
$\mathbf{h}_2=\sqrt{P_2}\mathbf{H}_2^H\mathbf{r}$. This result
finalizes the maximization with respect to $\mathbf{W}$.

\emph{\textbf{Step 2)} Maximization with respect to $\mathbf{s}$
and $\mathbf{r}$:}\\
From (16) we see that $\gamma$ is increasing both in
$\left\|\mathbf{h}_1\right\|$ and $\left\|\mathbf{h}_2\right\|$.
Therefore, for maximizing the SNR, we should maximize
$\left\|\mathbf{h}_1\right\|$ and $\left\|\mathbf{h}_2\right\|$,
subject to $\|\mathbf{s}\|=\|\mathbf{r}\|=1$. Considering the
definitions of $\mathbf{h}_1$ and $\mathbf{h}_2$, the optimal value
is achieved by letting $\mathbf{s}$ be the strongest right singular
vector of $\mathbf{H}_1$ and $\mathbf{r}$ be the strongest left
singular vector of $\mathbf{H}_2$. This concludes the maximization
in step 2.
\end{proof}

\begin{figure}%
\centering \subfigure[]{
\includegraphics[width=3.5in]{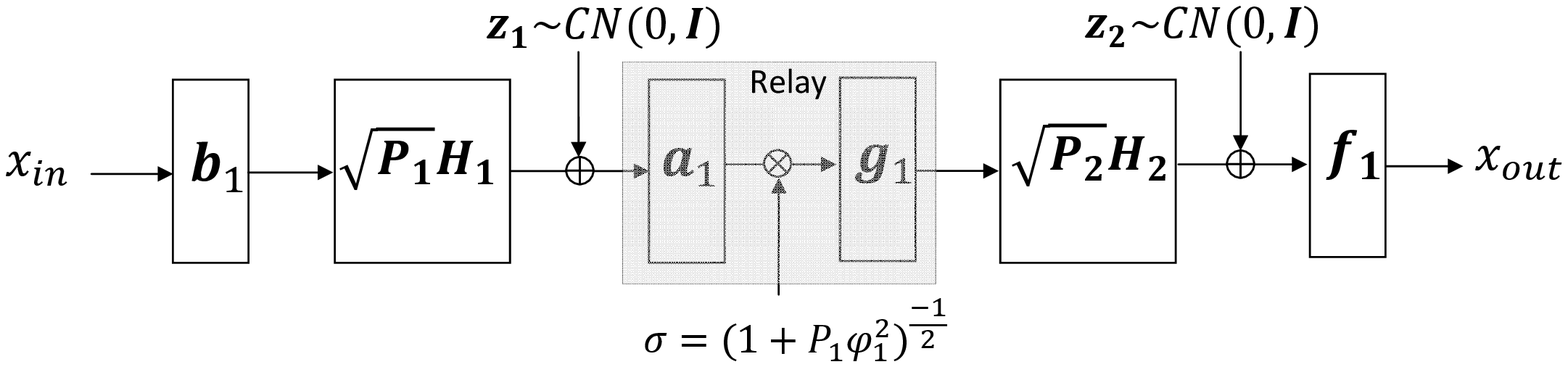}}\\
\subfigure[]{
\includegraphics[width=3.5in]{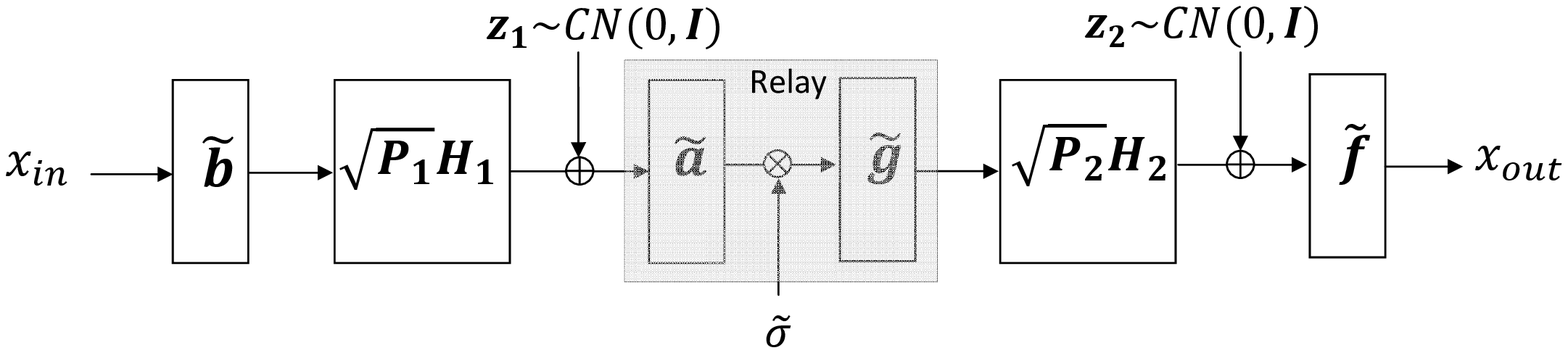}}
\caption[]{(a) Optimal unquantized scheme for MIMO AF without the
direct link, where
$\mathbf{H}_1=\mathbf{A}\mathbf{\Phi}\mathbf{B}^H$ and
$\mathbf{H}_2=\mathbf{F}\mathbf{\Psi}\mathbf{G}^H$. (b) Quantized
scheme for MIMO AF without the direct link.}
\end{figure}

Substituting the optimal solution, found in Theorem 1, in equation
(16) reveals the optimal SNR:
\begin{equation}
\gamma^\star=\frac{\gamma_1^\star\gamma_2^\star}{1+\gamma_1^\star+\gamma_2^\star},
\end{equation}
where
\begin{IEEEeqnarray}{lll}
\gamma_1^\star=\max_{\|\mathbf{s}\|=1}{P_1\|\mathbf{H}_1\mathbf{s}\|^2}=P_1\phi_1^2,\nonumber\\
\gamma_2^\star=\max_{\|\mathbf{s}\|=1}{P_2\|\mathbf{H}_2\mathbf{s}\|^2}=P_2\psi_1^2.
\end{IEEEeqnarray}

The optimal solution in Theorem 1 verifies the optimality of the
scheme in Fig. 3a, where the Tx and relay use the strongest right
singular vectors of the Tx-relay and relay-Rx channel matrices for
beamforming. Assuming that the relay knows $\mathbf{H}_1$ and the
receiver knows $\mathbf{H}_2$, the optimal structure can be achieved
if:
\begin{itemize}
\item The relay informs the transmitter of ${\mathbf{b}}_1$, the
strongest right singular vector of $\mathbf{H}_1$.
\item The receiver informs the relay of ${\mathbf{g}}_1$, the strongest
right singular vector of $\mathbf{H}_2$.
\end{itemize}
Considering this, we continue the problem in Subsection III.B by
characterizing the codebooks that should be used for quantizing the
optimal beamforming vectors.

\subsection{Quantization Scheme}
Fig. 3b presents a scheme which mimics the optimal scheme (Fig. 3a),
with the difference that the Tx and relay beamforming vectors belong
to certain codebooks with finite cardinality.

In Fig. 3b, the Tx beamforming vector $\tilde{\mathbf{b}}$ should
belong to a codebook $\mathbf{C}_1(N_1,\delta_1)$ shared between the
Tx and relay, and similarly, the relay beamforming vector
$\tilde{\mathbf{g}}$ should belong to a possibly different codebook
$\mathbf{C}_2(N_2,\delta_2)$, which is shared between the relay and
Rx. The relay and Rx use $\tilde{\mathbf{a}}$ and
$\tilde{\mathbf{f}}$ for receive beamforming, respectively. All
transmit/receive vectors $\tilde{\mathbf{a}}$, $\tilde{\mathbf{b}}$,
$\tilde{\mathbf{f}}$ and $\tilde{\mathbf{g}}$ are assumed to be of
unit norm, and
$\sigma=(1+P_1|\tilde{\mathbf{a}}^H\mathbf{H}_1\tilde{\mathbf{b}}|^2)^{-1/2}$
in order to satisfy the relay power constraint\footnote{The Tx power
constraint is automatically satisfied by assuming
$\|\tilde{\mathbf{b}}\|=1.$}. The received SNR of the quantized
scheme can be easily shown to be equal to:
\begin{equation}
\gamma=\frac{\gamma_1\gamma_2}{1+\gamma_1+\gamma_2},
\end{equation}
where
$\gamma_1=P_1\left|\tilde{\mathbf{a}}^H\mathbf{H}_1\tilde{\mathbf{b}}\right|^2$
and
$\gamma_2=P_2\left|\tilde{\mathbf{f}}^H\mathbf{H}_2\tilde{\mathbf{g}}\right|^2$
are the received SNRs of the Tx-relay and relay-Rx channels. As
$\gamma$ is increasing both in $\gamma_1$ and $\gamma_2$, we should
maximize these quantities to maximize the SNR of the quantized
scheme. This is accomplished, as in Section II, by letting
$\tilde{\mathbf{a}}$ and $\tilde{\mathbf{f}}$ to be matched to
$\mathbf{H}_1\tilde{\mathbf{b}}$ and
$\mathbf{H}_2\tilde{\mathbf{g}}$, and, choosing $\tilde{\mathbf{b}}$
and $\tilde{\mathbf{g}}$ based on: $\tilde{\mathbf{b}}=\arg
\max_{\mathbf{w}\in{\mathbf{C}_1}}{P_1\|\mathbf{H}_1{\mathbf{w}}\|^2}$
and
$\tilde{\mathbf{g}}=\arg\max_{\mathbf{w}\in{\mathbf{C}_2}}{P_2\|\mathbf{H}_2{\mathbf{w}}\|^2}$.
The corresponding received SNR values are
\begin{equation}\tilde{\gamma}_1=\max_{\mathbf{w}\in{\mathbf{C}_1}}{P_1\|\mathbf{H}_1{\mathbf{w}}\|^2},~~
\tilde{\gamma}_2=\max_{\mathbf{w}\in{\mathbf{C}_2}}{P_2\|\mathbf{H}_2{\mathbf{w}}\|^2},\end{equation}
and the maximum received SNR of the quantized scheme
$\tilde{\gamma}$ can be computed by substituting these quantities in
(19):\begin{equation}\tilde{\gamma}=\frac{\tilde{\gamma}_1\tilde{\gamma}_2}{1+\tilde{\gamma}_1+\tilde{\gamma}_2}.\end{equation}

In Appendix II.A, we use the distributions of the optimal
beamforming vectors $\mathbf{b}_1$ and $\mathbf{g}_1$ for Rayleigh
channels to compute the following upper bound for the total loss in
SNR caused by quantization.
\begin{IEEEeqnarray}{lll}
\mathrm{E}\{\gamma^\star\}-\mathrm{E}\{\tilde{\gamma}\}\leq
2mnP_1\left(1{-}N_1\left(\frac{\delta_1}{2}\right)^{2(m{-}1)}\left(1{-}\frac{\delta_1}{2}\right)\right)\nonumber\\
~~~~~~~~~~~~~+2nlP_2\left(1{-}N_2\left(\frac{\delta_2}{2}\right)^{2(n{-}1)}\left(1{-}\frac{\delta_2}{2}\right)\right),
\end{IEEEeqnarray}
This upper bound is decreasing in $\delta_1$ and $\delta_2$ for any
$m>1$ and $n>1$. Therefore, to minimize this upper bound, we should
maximize the minimum distances $\delta_1$ and $\delta_2$. This is
exactly the criterion used in Grassmannian codebook design and
proves the efficiency of these codebooks for quantizing the optimal
beamforming vectors. In Section V, we present simulation results
which compare the performance of the Grassmannian quantizers with
the optimal (unquantized) scheme and other possible quantization
schemes.
\section{MIMO Amplify and Forward Relay Channel with the Direct Link}
In this section the direct link is included in the system model
(Fig. 4). The optimal unquantized scheme is derived in Subsection
IV.A and the quantization scheme is presented in IV. B. Finally, in
IV.C we introduce a modified quantized scheme, which significantly
reduces the number of feedback bits with a negligible degradation in
the system performance.

\begin{figure}
\centering
\includegraphics[width=3.5in]{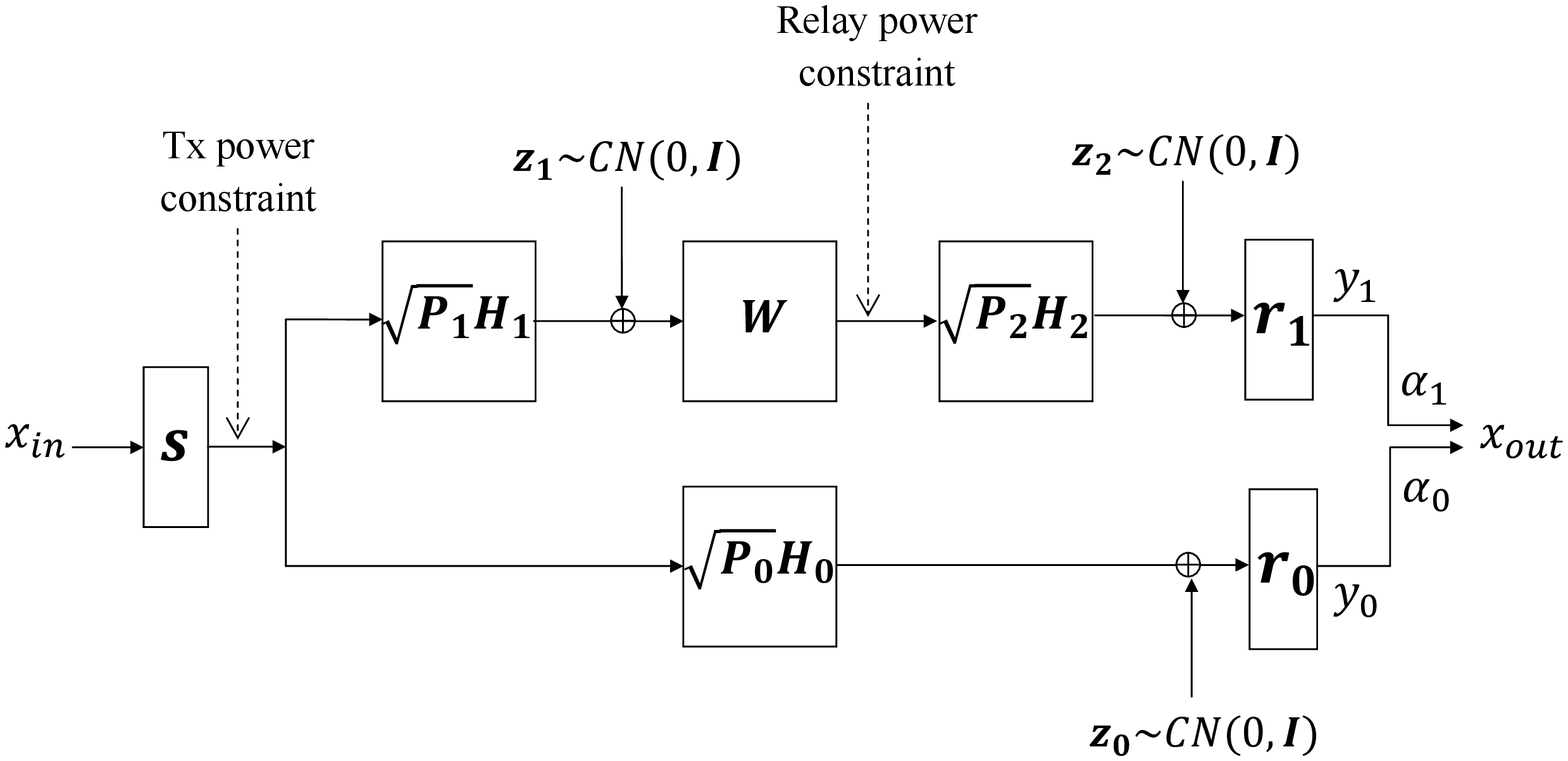}
\caption{Half-duplex MIMO AF relay channel model with direct link
between the transmitter and the receiver.}
\end{figure}

\subsection{Optimal Unquantized Scheme}
Consider the half-duplex MIMO-relay link in Fig. 4. At the first
time slot, the relay is silent and the Rx receives its symbol. At
the second time slot, the Tx is silent and the relay amplifies and
forwards its signal (received in the first time slot) to Rx. The
receiver has access to two received symbols $y_0$ and $y_1$
separated in time:
\begin{IEEEeqnarray}{l}
y_0=\sqrt{P_0}\mathbf{r}_0^H\mathbf{H}_0\mathbf{s}x+\mathbf{r}_0^H\mathbf{z}_0\nonumber\\
y_1=\sqrt{P_1P_2}\mathbf{r}_1^H\mathbf{H}_2\mathbf{W}\mathbf{H}_1\mathbf{s}x+\mathbf{r}_1^H\left(\sqrt{P_2}\mathbf{H}_2\mathbf{W}\mathbf{z}_1+\mathbf{z}_2\right).\nonumber
\end{IEEEeqnarray}
The receiver computes the linear MMSE combination of $y_0$ and $y_1$
to compute the output symbol $x_{out}$:
\[x_{out}=\alpha_0y_0+\alpha_1y_1.\]
By proper choice of $\alpha_0$ and $\alpha_1$ the output SNR
is\footnote{This is a result of the MMSE combination, or MRC after
scaling the noise levels of the symbols $y_0$ and $y_1$.}:
\begin{equation}
\gamma=\gamma_0+\gamma_{r},
\end{equation}
where $\gamma_0$ and $\gamma_r$ are the received SNR values of the
direct link and the Tx-relay-Rx link. Therefore, the total SNR is
maximized if the received SNRs of the direct and relay links are
maximized. The only common parameter in maximizing these two
quantities is the Tx beamforming vector $\mathbf{s}$.

By fixing $\mathbf{s}$ and following the same steps in Sections II
and III, the optimal values of other parameters can be easily
derived, as showed in Fig. 5a. In the first time slot, the relay and
the Rx should respectively match to $\mathbf{H}_1\mathbf{s}$ and
$\mathbf{H}_0\mathbf{s}$ at their inputs. In the second time slot,
the relay maps its normalized\footnote{To meet the relay power
constraint.} symbol to $\mathbf{g}_1$ the strongest right singular
vector of $\mathbf{H}_2$ and the receiver uses $\mathbf{f}_1$, the
strongest left singular vector of $\mathbf{H}_2$, for receive
beamforming. The corresponding received SNRs of the direct link and
relay link are:
\begin{IEEEeqnarray}{l}
\gamma_0=P_0\|\mathbf{H}_0\mathbf{s}\|^2\nonumber\\
\gamma_r=\frac{\gamma_1\gamma_2^\star}{1+\gamma_1+\gamma_2^\star},
\end{IEEEeqnarray}
where $\gamma_1=P_1\|\mathbf{H}_1\mathbf{s}\|^2$ and
$\gamma_2^\star$ is the maximum received SNR of the the relay-Rx
link:
$\gamma_2^\star=P_2\|\mathbf{H}_2\mathbf{g}_1\|^2=P_2\psi_1^2$. By
combining (23) and (24) the total received SNR is:
\[\gamma=\frac{P_1\|\mathbf{H}_1\mathbf{s}\|^2\gamma_2^\star}{1+P_1\|\mathbf{H}_1\mathbf{s}\|^2+\gamma_2^\star}+P_0\|\mathbf{H}_0\mathbf{s}\|^2,\]
and therefore, the optimal $\mathbf{s}$ can be expressed as:
\begin{equation}
\mathbf{s}^{\star}=\arg
\max_{\|\mathbf{s}\|=1}{\frac{\|\mathbf{H}_1\mathbf{s}\|^2}{\|\mathbf{H}_1\mathbf{s}\|^2+\lambda}+\mu\|\mathbf{H}_0\mathbf{s}\|^2},
\end{equation}
where $\lambda=\frac{1+\gamma_2^\star}{P_1}$ and
$\mu=\frac{P_0}{\gamma_2^\star}$. The corresponding total received
SNR is:
\begin{equation}\gamma^\star=\frac{\gamma_1^\star\gamma_2^\star}{1+\gamma_1^\star+\gamma_2^\star}+\gamma_0^\star,\end{equation}
where $\gamma_0^\star=P_0\|\mathbf{H}_0\mathbf{s}^\star\|^2$ and
$\gamma_1^\star=P_1\|\mathbf{H}_1\mathbf{s}^\star\|^2$.

The objective function of the problem in (25) has multiple local
maximum points and moreover, the global maximum point is not
unique\footnote{If $\mathbf{s}$ is a global maximum point, so is
$e^{j\theta}\mathbf{s}$, for any $\theta\in\mathds{R}$.}. This
problem does not appear to have an analytic solution and as a result
we use a numerical approach to perform this optimization, which will
be described in Section V.

\begin{figure}%
\centering \subfigure[]{
\includegraphics[width=3.5in]{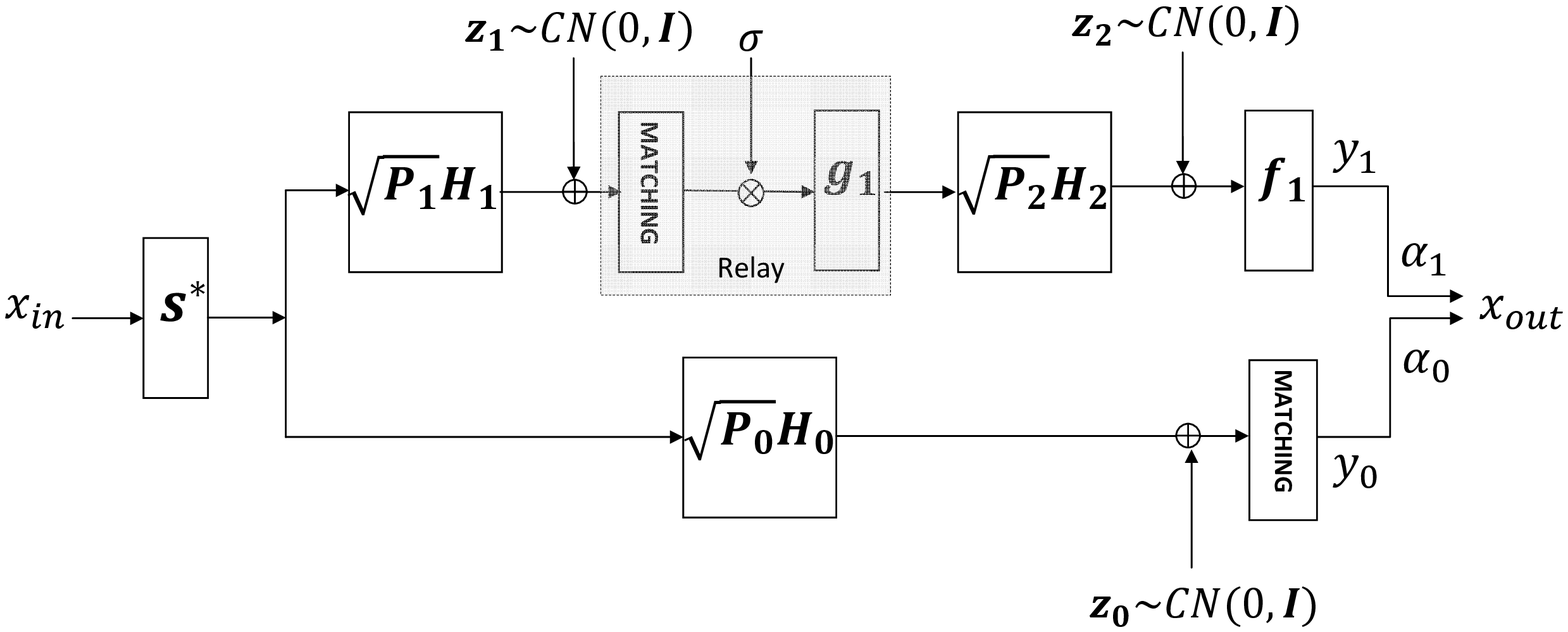}}\\
\subfigure[]{
\includegraphics[width=3.5in]{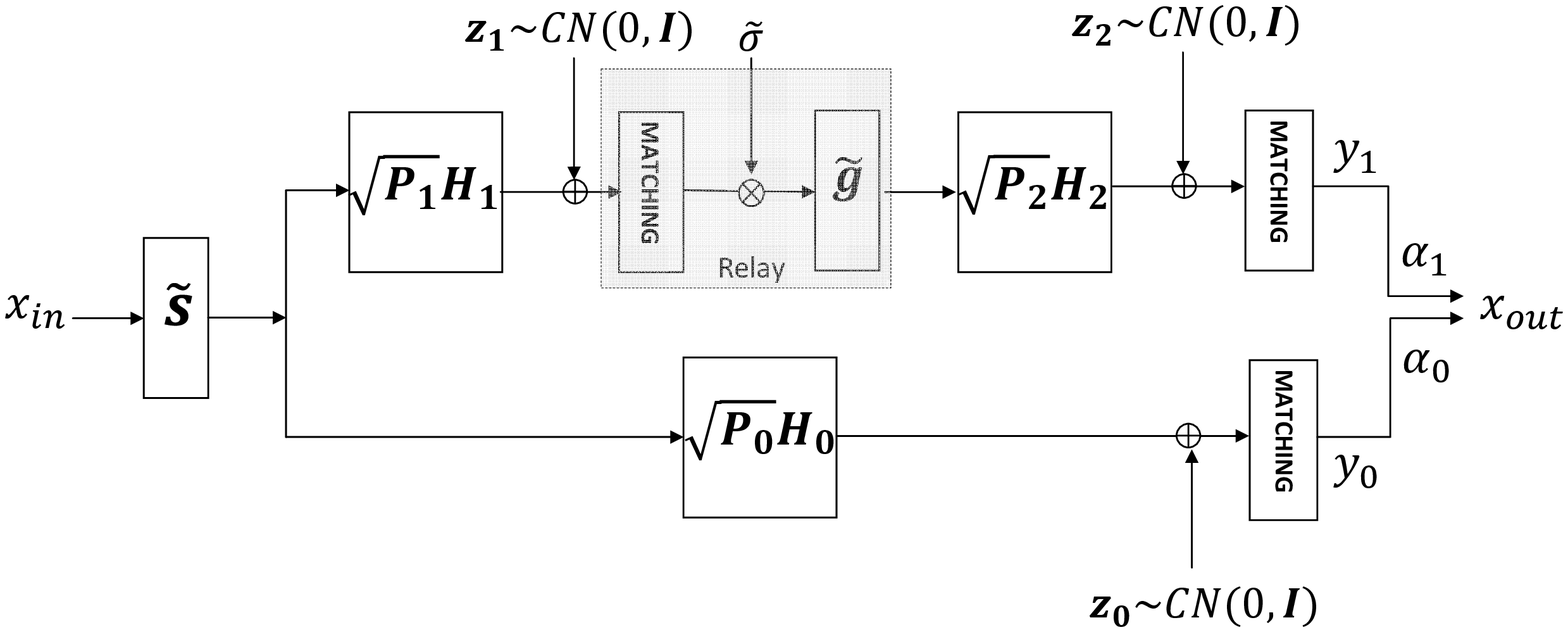}}
\caption[]{(a) Optimal unquantized scheme for MIMO AF with the
direct channel. In the first time slot, the relay and the Rx match
to $\mathbf{H}_1\mathbf{s}^\star$ and
$\mathbf{H}_0\mathbf{s}^\star$, respectively. (b) Quantized scheme
for MIMO AF with direct link. In the first time slot, the relay and
the Rx match to $\mathbf{H}_1\tilde{\mathbf{s}}$ and
$\mathbf{H}_0\tilde{\mathbf{s}}$, respectively. In the second time
slot, the relay matches to $\mathbf{H}_2\tilde{\mathbf{g}}$. When
$\tilde{\mathbf{s}}$ is replaced by $\tilde{\mathbf{s}}^\star$ in
(31), we will refer to this system as the ``properly quantized
scheme''.}
\end{figure}

Despite the fact that we do not have a closed form expression for
the solution of problem (25), we are still able to identify the
distribution of the solution for Rayleigh fading channels. The main
result of this section is the following theorem.

\begin{theorem}
For independent Rayleigh channel matrices $\mathbf{H}_0$ and
$\mathbf{H}_1$, the optimal Tx beamforming vector $\mathbf{s}^\star$
that maximizes the total received SNR (or equivalently the objective
function in (25)) is uniformly distributed on the unit sphere in
$\mathds{C}^m$, where $m$ is the number of Tx antennas.
\end{theorem}
\begin{proof}
See Appendix I.
\end{proof}

Note that if we had a single channel from the transmitter to the
receiver, the optimal Tx beamforming vector would be uniformly
distributed on the unit sphere in $\mathds{C}^m$ (see Section II).
Interestingly, Theorem 2 states that the optimal Tx beamforming
vector is still uniformly distributed on the unit sphere, when there
are two independent parallel channels from the transmitter to the
receiver. This is basically due to the independence of
$\mathbf{H}_0$ and $\mathbf{H}_1$, and the specific properties of
the Rayleigh channel matrices.

The result in Theorem 2 is used in Appendix II.B to derive an SNR
loss upper bound, similar to (2) and (22), which justifies use of
the Grassmannian codebook for quantizing the optimal Tx beamforming
vector $\mathbf{s}^{\star}$.
\subsection{Quantization Scheme}
Having identified the optimal scheme, we continue by considering the
quantization scheme in Fig. 5b. In the first time slot, the Tx uses
$\tilde{\mathbf{s}}$ for beamforming, and relay and Rx match their
receive vectors to $\mathbf{H}_1\tilde{\mathbf{s}}$ and
$\mathbf{H}_0\tilde{\mathbf{s}}$. In the second time slot, the relay
scales its symbol and uses $\tilde{\mathbf{g}}$ for beamforming and
Rx matches to $\mathbf{H}_2\tilde{\mathbf{g}}$. The Tx-Rx, Tx-relay,
relay-Rx and total received SNR values is given by:
\begin{IEEEeqnarray}{l}
{\gamma}_0=P_0\|\mathbf{H}_0\tilde{\mathbf{s}}\|^2,~~{\gamma}_1=P_1\|\mathbf{H}_1\tilde{\mathbf{s}}\|^2,~~{\gamma}_2=P_2\|\mathbf{H}_2\tilde{\mathbf{g}}\|^2\nonumber\\
\gamma=\frac{\gamma_1\gamma_2}{1+\gamma_1+\gamma_2}+{\gamma}_0.
\end{IEEEeqnarray}
We need to maximize (27) with respect to the Tx and relay
beamforming vectors $\tilde{\mathbf{s}}$ and $\tilde{\mathbf{g}}$,
which belong to certain codebooks with finite cardinalities. As in
Section III, we assume that the codebooks
$\mathbf{C}_1(N_1,\delta_1)$ and $\mathbf{C}_2(N_2,\delta_2)$ are
shared between Tx-relay and relay-Rx, respectively. Clearly,
$\tilde{\mathbf{g}}$ should be chosen to maximize ${\gamma}_2$:
\begin{equation}
\tilde{\mathbf{g}}=\arg\max_{\mathbf{w}\in
\mathbf{C}_2}{P_2\|\mathbf{H}_2{\mathbf{w}}\|^2}.
\end{equation}
The corresponding relay-Rx received SNR is:
$\tilde{\gamma}_2=\max_{\mathbf{w}\in
\mathbf{C}_2}{P_2\|\mathbf{H}_2{\mathbf{w}}\|^2}$.

For choosing the proper $\tilde{\mathbf{s}}$, we need to know both
$\mathbf{H}_0$ and $\mathbf{H}_1$. We continue the problem here by
assuming that the relay knows $\mathbf{H}_0$ in addition to its
channel $\mathbf{H}_1$. This assumption will be relaxed in IV.B.2.

\subsubsection{Complete Knowledge of $\mathbf{H}_0$ at the Relay}
If the relay knows both $\mathbf{H}_0$ and $\mathbf{H}_1$, then
based on (27) the best vector $\tilde{\mathbf{s}}$ should be chosen
as follows:
\begin{equation}
\tilde{\mathbf{s}}=\arg\max_{\mathbf{w}\in\mathbf{C}_1}{
\frac{\|\mathbf{H}_1\mathbf{w}\|^2}{\|\mathbf{H}_1\mathbf{w}\|^2+\tilde{\lambda}}+\tilde{\mu}\|\mathbf{H}_0\mathbf{w}\|^2},
\end{equation}
where $\tilde{\lambda}=\frac{1+\tilde{\gamma}_2}{P_1}$ and
$\tilde{\mu}=\frac{P_0}{\tilde{\gamma}_2}$. The maximum total
received SNR of the quantized scheme $\tilde{\gamma}$ can be
computed by substituting (28) and (29) in (27).

In Appendix II.B, we use the distribution of $\mathbf{s}^{\star}$,
given in Theorem 2, to prove the following bound on the SNR loss
caused by quantization.
\begin{IEEEeqnarray}{lll}
\mathrm{E}\{\gamma^\star\}-\mathrm{E}\{\tilde{\gamma}\}\nonumber\\
~~~\leq2\left(mlP_0+mnP_1\right)\left(1{-}N_1\left(\frac{\delta_1}{2}\right)^{2(m{-}1)}\left(1{-}\frac{\delta_1}{2}\right)\right)\nonumber\\
~~~~~~~~~~~~+2nlP_2\left(1{-}N_2\left(\frac{\delta_2}{2}\right)^{2(n{-}1)}\left(1{-}\frac{\delta_2}{2}\right)\right).
\end{IEEEeqnarray}
This upper bound is decreasing in $\delta_1=\delta(\mathbf{C}_1)$
and $\delta_2=\delta(\mathbf{C}_2)$ for any $m,n>1$ and justifies
the use of Grassmannian codebooks, $\mathbf{C}_1$  and
$\mathbf{C}_2$, for quantizing the optimal Tx and relay beamforming
vectors, $\mathbf{s}^\star$ and $\mathbf{g}_1$.

\subsubsection{Partial Knowledge of $\mathbf{H}_0$ at the Relay}
As mentioned earlier, the computations in IV.B.1 are based on the
assumption that the relay knows $\mathbf{H}_0$ completely. In
reality, however, the Rx needs to quantize $\mathbf{H}_0$ and send
it to the relay. We should note that, the only way that
$\mathbf{H}_0$ contributes to the problem in (28) is through the
term $\|\mathbf{H}_0\mathbf{w}\|^2$, which can be expanded as
follows:
$\|\mathbf{H}_0\mathbf{w}\|^2=\sum_{i=1}^{R_0}{\nu_i^2|\mathbf{e}_i^H\mathbf{w}|^2}$,
where $\nu_i$'s and $\mathbf{e}_i$'s are the singular values and
right singular vectors of $\mathbf{H}_0$ and
$R_0=\texttt{rank}({\mathbf{H}_0})$. Therefore, the relay only needs
to know the singular values and the right singular vectors of the
direct link channel. Since our focus in this paper is on the vector
quantization feedback schemes, we assume that the relay knows the
singular values completely but has only access to the quantized
versions of the singular vectors.

For quantizing the singular vectors, the Rx and the relay share a
codebook $\mathbf{C}_0(N_0,\delta_0)$, which is possibly different
from $\mathbf{C}_2$ (used for determining $\tilde{\mathbf{g}}$). We
assume that the Rx quantizes each vector $\mathbf{e}_i$ to a vector
$\tilde{\mathbf{e}}_i\in\mathbf{C}_0$ that is closest to
$\mathbf{e}_i$.
\[\tilde{\mathbf{e}}_i=\arg\min_{\mathbf{w}\in\mathbf{C}_0}{d(\mathbf{w},\mathbf{e}_i)}.\]
Having $\nu_i$'s and $\tilde{\mathbf{e}}_i$'s at the relay, the
problem of finding the Tx beamforming vector
$\tilde{\mathbf{s}}^\star$ can be reformulated as\footnote{Here, we
have used the notation $\tilde{\mathbf{s}}^\star$ to distinguish
this vector form the vector $\tilde{\mathbf{s}}$ in (28), where we
were assuming that the relay knows $\mathbf{H}_0$ completely.}:
\begin{equation}
\tilde{\mathbf{s}}^\star=\arg\max_{\mathbf{w}\in\mathbf{C}_1}{
\frac{\|\mathbf{H}_1\mathbf{w}\|^2}{\|\mathbf{H}_1\mathbf{w}\|^2+\tilde{\lambda}}+\tilde{\mu}{\sum_{i=1}^{R_0}{\nu_i^2|\tilde{\mathbf{e}}_i^H\mathbf{w}|^2}},}
\end{equation}
where $\tilde{\lambda}=\frac{1+\tilde{\gamma}_2}{P_1}$,
$\tilde{\mu}=\frac{P_0}{\tilde{\gamma}_2}$, and
$\tilde{\gamma}_2=\max_{\mathbf{w}\in
\mathbf{C}_2}{P_2\|\mathbf{H}_2{\mathbf{w}}\|^2}$. The total
received SNR $\tilde{\gamma}^\star$ can be computed by substituting
(28) and (31) in (27). Finally, the loss in the received SNR can be
bounded as follows (see Appendix II.C).
\begin{IEEEeqnarray}{lll}
\mathrm{E}\{\gamma^\star\}-\mathrm{E}\{\tilde{\gamma}^\star\}\nonumber\\
~\leq2\left(mlP_0+mnP_1\right)\left(1{-}N_1\left(\frac{\delta_1}{2}\right)^{2(m{-}1)}\left(1{-}\frac{\delta_1}{2}\right)\right)\nonumber\\
~~~~~~~~~~+2nlP_2\left(1{-}N_2\left(\frac{\delta_2}{2}\right)^{2(n{-}1)}\left(1{-}\frac{\delta_2}{2}\right)\right)\nonumber\\
~~~~~~~~~~+4mlP_0\left(1{-}N_0\left(\frac{\delta_0}{2}\right)^{2(m{-}1)}\left(1{-}\frac{\delta_0}{2}\right)\right).
\end{IEEEeqnarray}
The upper bound in (32) is decreasing in
$\delta_0=\delta(\mathbf{C}_0)$ for any $m>1$. This justifies use of
the Grassmannian codebook to quantize the singular vectors of
$\mathbf{H}_0$, since it has the maximum minimum distance
$\delta_0$. The same conclusion holds for $\mathbf{C}_1$ and
$\mathbf{C}_2$, since the upper bound in (32) is decreasing in
$\delta_1$ and $\delta_2$ for any $m,n>1$.

To summarize the results, all three codebooks $\mathbf{C}_0$,
$\mathbf{C}_1$ and $\mathbf{C}_2$ need to be Grassmannian codebooks
to minimize the upper bound of the loss in the total received SNR.
We refer to the scheme, determined by (31), as the ``properly
quantized scheme''. In the following we outline the steps in
determining the beamforming vectors of the ``properly quantized
scheme'' (Fig. 5b).
\begin{enumerate}
\item The Rx uses a Grassmannian codebook $\mathbf{C}_2$,
shared between the Rx and the relay, to quantize $\mathbf{g}$, the
strongest right singular vector of the relay-Rx channel
$\mathbf{H}_2$. The label of the quantized vector is sent to the
relay. The relay uses this vector for its beamforming in the second
time slot. The Rx also sends the SNR value $\tilde{\gamma}_2$ to the
relay. This will be used in step 3.
\item The Rx quantizes the right singular vectors of the Tx-Rx channel
using a Grassmannian codebook $\mathbf{C}_0$, which is shared
between the Rx and the relay. The labels of the quantized vectors
and the singular values $\nu_i$'s are sent to the relay.
\item The relay forms the objective function in (31) and maximizes it
over the Grassmannian codebook $\mathbf{C}_1$, which is shared
between the Tx and the relay. The relay sends the label of the
maximizing vector to the Tx. The transmitter uses this vector for
its beamforming in the first time slot.
\end{enumerate}

Before concluding Section IV, we introduce a modified scheme which
performs very close to the ``properly quantized scheme'' but
requires fewer number of feedback bits.

\subsection{Modified Quantized Scheme}
Consider the problem of determining the Tx beamforming vector for
the quantized scheme in Fig. 5b (see equation (31)). There are two
links between the transmitter and the receiver; the direct (Tx-Rx)
link and the Tx-relay-Rx link, which we refer to as the relay link.
If the direct link is much weaker than the relay link and can be
ignored safely, our problem reduces to the problem in Section III
and the relay does not need to know anything about the direct link
channel $\mathbf{H}_0$. On the other hand, if the relay link is very
weak and can be ignored, the only thing that we need to know about
$\mathbf{H}_0$ is its strongest right singular vector. Therefore, in
both of these extreme cases we do not need to have any knowledge of
$\mathbf{H}_0$ other than its strongest right singular vector. Based
on this intuition, we propose a new scheme, referred to as the
``modified quantized scheme'', in which the Rx only quantizes the
strongest right singular vector of $\mathbf{H}_0$ and sends the
corresponding label (and the largest singular value $\nu_1$) to the
relay. The relay then determines the proper Tx beamforming vector by
forming the following problem.
\begin{equation}
\tilde{\mathbf{s}}_{modified}=\arg\max_{\mathbf{w}\in\mathbf{C}_1}{
\frac{\|\mathbf{H}_1\mathbf{w}\|^2}{\|\mathbf{H}_1\mathbf{w}\|^2+\tilde{\lambda}}+\tilde{\mu}\nu_1^2|\tilde{\mathbf{e}}_1^H\mathbf{w}|^2},
\end{equation}
where $\tilde{\lambda}$ and $\tilde{\mu}$ have the same definitions
as in (31).

The ``modified quantized scheme'' requires much fewer number of
bits, since it only quantizes one singular vector (see step 2 for
the properly quantized scheme). Our simulation results show that the
``modified quantized scheme'' performs very close to the ``properly
quantized scheme'', as we will see in Section V.

\section{Simulation Results}
In this section, we provide simulation results for the scenarios
discussed in the Sections III and IV. The results are divided into
two subsections. In V.A the direct link between the transmitter and
the receiver is ignored, as in Section III (see Fig. 2). In V.B, the
simulation results are presented for the case where the direct link
is present in the model (Fig. 4).

The general setup for the simulations is as follows. The input
symbols belong to a BPSK constellation with unit power. The entries
of the channel matrices, which model the i.i.d Rayleigh fading
channels, are generated independently according to
$\mathcal{CN}(0,1)$. To model quasi-static fading channels, the
simulation time is divided to $20,000$ coherence intervals, each
consisting of $200$ symbols. The channels are assumed to be constant
over each coherence interval and to be independent from one interval
to the other. The simulation results compare different (quantized
and unquantized) schemes from the bit-error-rate (BER) point of
view.

\subsection{MIMO AF Relay Channel without the Direct Link}

In this section, the direct link is not considered in the simulation
model (Fig. 2). All of the stations (Tx, relay and Rx) are assumed
to have two antennas ($m=n=l=2$). The relay-Rx link SNR is fixed at
$P_2=8$dB and the BER values have been recorded for different values
of the Tx-relay link SNR $P_1$. For the quantization purposes, the
Tx and relay share a codebook $\mathbf{C}_1$ of size $N_1$.
Similarly, the relay and Rx share a codebook $\mathbf{C}_2$ of size
$N_2$.

\begin{figure}
\centering
\includegraphics[width=3.8in]{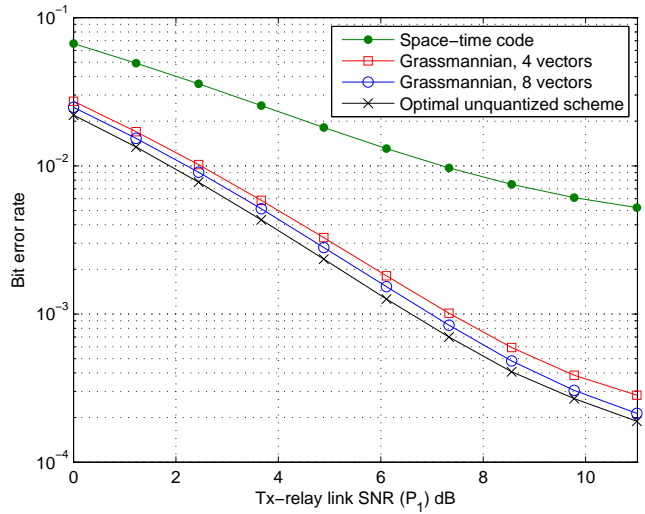}
\caption{Comparison of the performance of Grassmannian quantization
scheme with the optimal (unquantized) scheme and the Alamouti
space-time coding. The relay-Rx link SNR is fixed at $8$dB.}
\end{figure}

\begin{figure}
\centering
\includegraphics[width=3.8in]{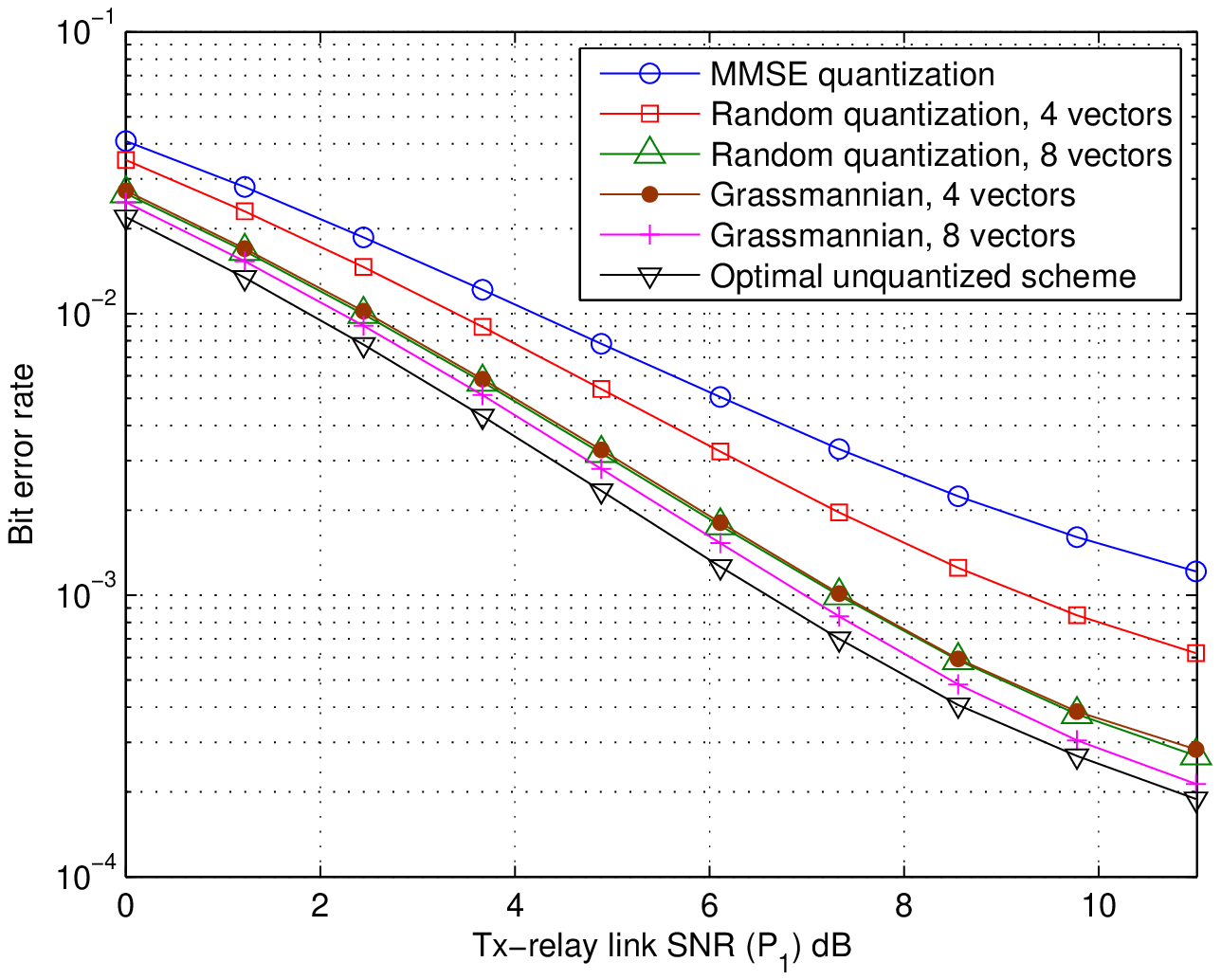}
\caption{Comparison of the performance of the Grassmannian quantizer
with the MMSE and random quantizers. The relay-Rx link SNR is fixed
at $8$dB.}
\end{figure}

Fig. 6 compares the performance of the ``optimal unquantized
scheme'' (Fig. 3a) with the performance of the Grassmannian
codebooks $\mathbf{C}_1$ and $\mathbf{C}_2$ of sizes $N_1=N_2=4$ or
$8$. The Grassmannian codebooks are adopted from [5]. The total
number of the feedback bits used by the Grassmannian quantizer is
$\log_2{N_1}+\log_2{N_2}$ which equals $4$ or $6$ bits for
$N_1=N_2=4$ or $8$. As Fig. 6 shows, we can get very close to the
optimal scheme with only a few number of bits per each coherence
interval. We have also simulated the performance of the Alamouti
code, to show the high power gain that can be achieved by using the
Grassmannian codebooks compared to space-time codes\footnote{In the
implementation of the Alamouti code, we have assumed that the relay
does not perform any decoding on its received symbols, to comply
with the amplify-and-forward assumption. The relay decomposes the
symbols coded by the Almouti code, and performs another Alamouti
coding on the decomposed symbols and sends the scaled symbols
through the relay-Rx channel.}.

In Fig. 7 we compare the performance of the Grassmannian quantizers
with other quantization schemes. For the MMSE quantization scheme,
the Rx and relay quantize every entry of the channel matrices
$\mathbf{H}_2$ and $\mathbf{H}_1$ according to the MMSE criterion
and send the quantized channel matrices to the relay and Tx,
respectively. The Tx and the relay perform singular value
decomposition on these quantized matrices and use the corresponding
strongest right singular vectors for beamforming. We have assumed
that the quantizer uses two bits to quantize each channel entry,
i.e., one bit for each of the real and imaginary parts. For
$m=n=l=2$ this results in $2(mn+nl)=16$ bits which should be
compared to the small number of feedback bits in the Grassmannian
scheme.

Fig. 7 also compares the Grassmannian quantizer with the random
quantization scheme. The random quantizer uses a set of randomly
selected vectors on the unit sphere as its quantization codebook.
The performance of the random scheme has been averaged over ten such
codebooks. As Fig. 7 shows, the Grassmannian scheme shows
considerable gain as compared with the random quantizer. However,
this gain decreases as the codebook sizes are increased from $4$ to
$8$. The main advantage of the random codebooks is that they are
easy to generate as compared with the Grassmannian codebooks.

\subsection{MIMO AF Relay Channel with the Direct Link}
In this section, we simulate the system model in Section IV, where
the direct link has been included in the analysis. All the stations
are equipped with three antennas ($m=n=l=3$).

Fig. 8 compares the ``optimal unquantized scheme'' (Fig. 5a) with
some other unquantized schemes. The Tx-relay and relay-Rx link SNR's
are fixed at $P_1=P_2=2$dB and the BER values are recorded for
different values of the direct link SNR $P_0$. For the optimal
scheme, we use the gradient descent method for determining the Tx
beamforming vector from (25). The constraint $\|\mathbf{s}\|=1$ is
eliminated by the change of variable
$\mathbf{s}=\frac{\mathbf{u}}{\|\mathbf{u}\|}$.

The curve marked by $\triangledown$ shows the performance of the
scheme that ignores the direct link in determining the Tx
beamforming vector. For this scheme, the Tx beamforming vector is
always set to the strongest right singular vector of the Tx-relay
channel. As expected, the performance of this scheme diverges from
the optimal scheme as the direct link gets stronger. The next curve,
marked by $\square$, shows the performance of the scheme which
considers only the stronger link for determining the Tx beamforming
vector. In this scheme, the Tx switches between the strongest right
singular vectors of the Tx-relay and Tx-Rx links depending on their
received SNR values. The last scheme, called the ``modified
unquantized scheme'', has the same structure as the ``optimal
unquantized scheme'' with the difference that the relay only
considers the strongest singular value and singular vector of
$\mathbf{H}_0$ in formulating the problem of determining the Tx
beamforming vector. This problem is exactly the same as the problem
(25), used by the optimal scheme, except that
$\|\mathbf{H}_0\mathbf{s}\|^2$ is replaced by
$\nu_1^2|\mathbf{e}_1^H\mathbf{s}|^2$, where $\nu_1$ and
$\mathbf{e}_1$ are the strongest singular value and right singular
vector of $\mathbf{H}_0$. In Appendix III, we show that the average
SNR loss of this scheme with respect to the optimal scheme is at
most $1.24$dB for the system with $m=n=l=3$ antennas. As the
simulation results in Fig. 8 verify, the modified unquantized scheme
performs very close to the optimal scheme. This unquantized scheme
is the basis for a quantization scheme that has been referred to as
the ``modified quantized scheme'' in Section IV (see (33)).

\begin{figure}
\centering
\includegraphics[width=3.8in]{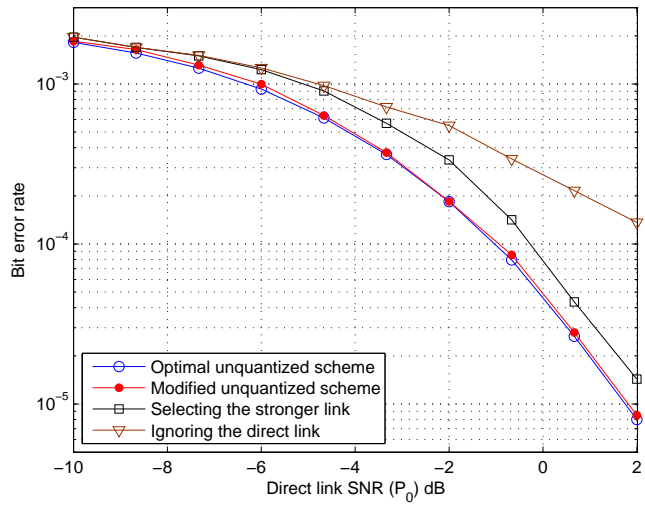}
\caption{Comparison of the optimal unquantized scheme with other
unquantized schemes. The Tx-relay and relay-Rx link SNRs are fixed
at $P_1=P_2=2$dB.}
\end{figure}

In the next two simulation setups, we study the performance of the
quantized schemes. As discussed in Section IV, the scheme consists
of three codebooks $\mathbf{C}_0$, $\mathbf{C}_1$ and $\mathbf{C}_2$
of sizes $N_0$, $N_1$ and $N_2$. The codebook $\mathbf{C}_0$ is used
for quantization of the direct link channel $\mathbf{H}_0$. The
codebook $\mathbf{C}_2$ is used to determine the relay beamforming
vector in the second time slot. The codebook $\mathbf{C}_1$
determines the Tx beamforming vector in the first time slot. Fig. 9
shows the performance of the ``properly quantized scheme'' with
Grassmannian codebooks of sizes $N_1=N_2=N_3=8,16$ (see the three
steps for properly quantized scheme in Section IV). The Tx-relay and
relay-Rx link SNRs are fixed at $P_1=P_2=2$dB and the BER values
have been recorded for different values of the direct link SNR
$P_0$. The Grassmannian codebooks are adopted from [14].

The figure also shows the performance of the Grassmannian codebooks
with ``modified quantized scheme'' (see (33)). This scheme shows a
negligible performance degradation with respect to the ``properly
quantized scheme'', but requires fewer number of feedback bits. As
an example, we compare the total number of bits required by the
properly quantized and the modified quantized scheme. For
quantization of the scalar values, we assume a hypothetical
quantizer which requires $b$ bits for quantizing a scalar quantity.
Recall the three steps of the properly quantized scheme in Section
IV. For step one, we need $\log_2(N_2)$ bits for quantizing
$\mathbf{g}$ and $b$ bits for quantizing $\tilde{\gamma}_2$. In step
two, we need $R_0(\log_2(N_0)+b)$ for the ``properly quantized
scheme'' and $\log_2(N_0)+b$ bits for the ``modified quantized
scheme'', where $R_0=\texttt{rank}(\mathbf{H}_0)$. Finally, for the
third step, we need $\log_2(N_1)$ bits for quantizing the Tx
beamforming vector. Therefore, we need a total of
$(1+R_0)b+\log_2(N_0^{R_0}N_1N_2)$ bits for the ``properly quantized
scheme'' and $2b+\log_2(N_0N_1N_2)$ bits for the ``modified
quantized scheme''. Table I compares these values for
$N=N_0=N_1=N_2=8,~16$, and $m=n=l=3$. Here we have assumed a full
rank channel matrix $\mathbf{H}_0$.


\begin{table}
\caption{Comparison of the Number of the Feedback Bits for Different
Quantization Schemes}
\begin{center}
\begin{tabular}{|c|c|c|}
\hline Scheme & \multicolumn{2}{|c|}{Number of feedback bits}\\
\cline{2-3} &$~~N=8~~$ & $N{=}16$\\ \hline \hline
Properly quantized & $15+4b$ & $20+4b$ \\
\hline
Modified quantized & $9+2b$ & $12+2b$\\
\hline
MMSE & \multicolumn{2}{|c|}{$54+b$}\\
\hline
\end{tabular}
\end{center}
\end{table}

Fig. 9 also shows the performance of the MMSE quantizer. This
quantizer requires $2(mn+ml+ln)$ bits for quantizing the channel
matrices and $b$ bits for quantizing $\tilde{\gamma}_2$.

\begin{figure}
\centering
\includegraphics[width=3.8in]{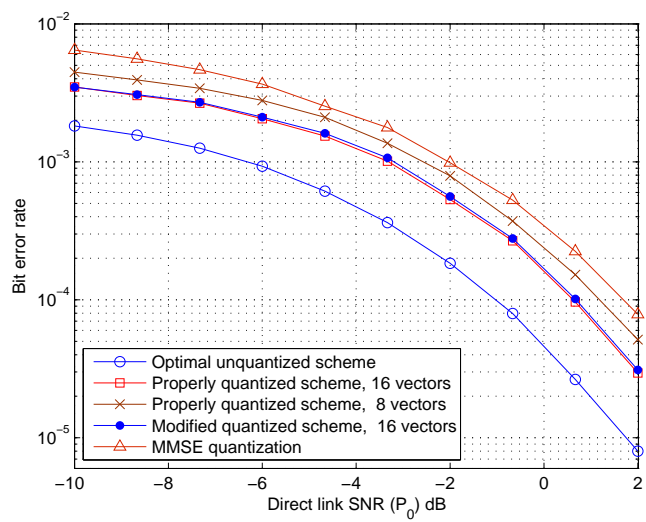}
\caption{Comparison of the properly quantized scheme with modified
quantized and MMSE quantization schemes. The Tx-relay and relay-Rx
link SNRs are fixed at $P_1=P_2=2$dB.}
\end{figure}

Fig. 10 compares the performance of the same schemes of Fig. 9 in a
different scenario. For this figure, the direct link and relay-Rx
link SNR are fixed at $P_0=-4$dB and $P_2=2$dB. The BER values have
been recorded for different values of the Tx-relay link SNR $P_1$.
Once again, we see that the performance of the ``modified quantized
scheme'' is very close to the ``properly quantized scheme''.

\begin{figure}
\centering
\includegraphics[width=3.8in]{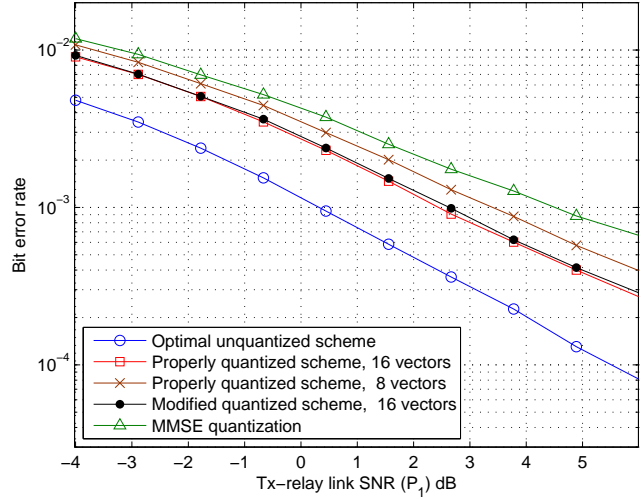}
\caption{Comparison of the properly quantized scheme with modified
quantized and MMSE quantization schemes. The direct link and
relay-Rx link SNR are fixed at $P_0=-4$dB and $P_2=2$dB.}
\end{figure}

\section{Conclusion}
In this paper, we derived the optimal (unquantized) Tx/Rx
beamforming vectors and the optimal relay weighting matrix to
maximize the total received SNR of MIMO AF relay channel both with
and without the direct Tx-Rx link. We showed that the Grassmannian
codebooks are appropriate choices for the quantization codebooks in
the quantized scheme. We proposed a modified quantized scheme which
performs very close to this quantized scheme and requires
considerably fewer number of feedback bits. Finally, the analytical
results were verified by comparing the performance of the
unquantized and quantized schemes under different scenarios.

\appendices \numberwithin{equation}{section}

\section{The Distribution of the Optimal Beamforming Vector
$\mathbf{s}^\star$}

In this appendix, we show that there exists a solution
$\mathbf{s}^\star$ to the problem (25) that is uniformly distributed
on the unit sphere in $\mathds{C}^m$, where $m$ is the number of Tx
antennas.

The problem (25) is repeated here:
\begin{equation}
\mathbf{s}^{\star}=\arg
\max_{\|\mathbf{s}\|=1}{\frac{\|\mathbf{H}_1\mathbf{s}\|^2}{\|\mathbf{H}_1\mathbf{s}\|^2+\lambda}+\mu\|\mathbf{H}_0\mathbf{s}\|^2},
\end{equation}
Consider $\mathbf{H}_0=\mathbf{U}_0\mathbf{\Sigma}_0\mathbf{V}_0^H$
and $\mathbf{H}_1=\mathbf{U}_1\mathbf{\Sigma}_1\mathbf{V}_1^H$ as
the SVD of $\mathbf{H}_0$ and $\mathbf{H}_1$. Clearly:
$\|\mathbf{H}_0\mathbf{s}\|=\left\|\mathbf{\Sigma}_0\mathbf{V}_0^H\mathbf{s}\right\|$
and
$\left\|\mathbf{H}_1\mathbf{s}\right\|=\left\|\mathbf{\Sigma}_1\mathbf{V}_1^H\mathbf{s}\right\|$,
since $\mathbf{U}_0$ and $\mathbf{U}_1$ are unitary matrices.

It is easy to check that
$\mathbf{s}^\star=\mathbf{V}_0\eta(\mathbf{\Sigma}_0,\mathbf{\Sigma}_1,\mathbf{V}_1^H\mathbf{V}_0)$
is a solution to (I.1), where the function $\eta(\cdot,\cdot,\cdot)$
is defined to be a solution to the following problem:
\begin{IEEEeqnarray}{lll}
\eta(\mathbf{\Sigma}_0,&\mathbf{\Sigma}_1&,\mathbf{V}_1^H\mathbf{V}_0)\nonumber\\
&\stackrel{def}{=}&\arg\max_{\|\mathbf{t}\|=1}{
\frac{\left\|\mathbf{\Sigma}_1\mathbf{V}_1^H\mathbf{V}_0\mathbf{t}\right\|^2}{\left\|\mathbf{\Sigma}_1\mathbf{V}_1^H\mathbf{V}_0\mathbf{t}\right\|^2+\lambda}+
\mu\left\|\mathbf{\Sigma}_0\mathbf{t}\right\|^2}.
\end{IEEEeqnarray}
If we fix $\mathbf{\Sigma}_0$ and $\mathbf{\Sigma}_1$, the solution
$\mathbf{s}^\star$, identified above, can be expressed as a function
of $\mathbf{V}_0$ and $\mathbf{V}_1$:
\begin{equation}
\mathbf{s}^\star=\zeta_{_{\mathbf{\Sigma}_0,\mathbf{\Sigma}_1}}(\mathbf{V}_0,\mathbf{V}_1)\stackrel{def}{=}\mathbf{V}_0\eta(\mathbf{\Sigma}_0,\mathbf{\Sigma}_1,\mathbf{V}_1^H\mathbf{V}_0).
\end{equation}

Now, for any unitary matrix $\mathbf{Q}$, we have the following from
(I.3).
\begin{equation}
\zeta_{_{\mathbf{\Sigma}_0,\mathbf{\Sigma}_1}}(\mathbf{Q}\mathbf{V}_0,\mathbf{Q}\mathbf{V}_1)=\mathbf{Q}\zeta_{_{\mathbf{\Sigma}_0,\mathbf{\Sigma}_1}}(\mathbf{V}_0,\mathbf{V}_1)
=\mathbf{Q}\mathbf{s}^\star.\nonumber
\end{equation}
For a Rayleigh channel matrix $\mathbf{H}_0$, we know the the random
matrix $\mathbf{V}_0$ is independent of $\mathbf{\Sigma}_0$ and its
distribution does not change by pre-multiplication by a unitary
matrix $\mathbf{Q}$. The same argument holds for $\mathbf{H}_1$,
$\mathbf{V}_1$ and $\mathbf{\Sigma}_1$. Therefore, conditioned on
$\mathbf{\Sigma}_0$ and $\mathbf{\Sigma}_1$, the matrix
$\mathbf{Q}\mathbf{V}_0$ has the same distribution as
$\mathbf{V}_0$, and similarly $\mathbf{Q}\mathbf{V}_1$ has the same
distribution as $\mathbf{V}_1$. Since the Tx-Rx and Tx-relay
channels are assumed to be independent, $\mathbf{V}_0$ and
$\mathbf{V}_1$ are also independent, and therefore the joint
distribution of $(\mathbf{V}_0,\mathbf{V}_1)$ is also the same as
the joint distribution of
$(\mathbf{Q}\mathbf{V}_0,\mathbf{Q}\mathbf{V}_1)$. Hence, any
arbitrary function of these pairs will have the same distribution.
By applying this to the function
$\zeta_{_{\mathbf{\Sigma}_0,\mathbf{\Sigma}_1}}(\cdot)$, we conclude
that
$\mathbf{s}^\star=\zeta_{_{\mathbf{\Sigma}_0,\mathbf{\Sigma}_1}}(\mathbf{V}_0,\mathbf{V}_1)$
and $\mathbf{Q}\mathbf{s}^\star=
\zeta_{_{\mathbf{\Sigma}_0,\mathbf{\Sigma}_1}}(\mathbf{Q}\mathbf{V}_0,\mathbf{Q}\mathbf{V}_1)$
have the same distribution. Since this it true for any unitary
matrix $\mathbf{Q}$, we conclude that $\mathbf{s}^\star$ is
uniformly distributed on the complex unit sphere, conditioned on
$\mathbf{\Sigma}_0$ and $\mathbf{\Sigma}_1$.

Note that if the conditional distribution of $\mathbf{s}^\star$ is
uniform, its unconditional distribution is also uniform. Moreover,
the random vector $\mathbf{s}^\star$ is independent of the random
matrices $\mathbf{\Sigma}_0$ and $\mathbf{\Sigma}_1$, since its
conditional and unconditional distributions are the same.

\section{Proof of SNR Loss Upper Bounds}

In this appendix, we prove the SNR loss upper bounds of (22), (30)
and (32) in three separate sections. We will first prove the
following lemmas, which are frequently used in these sections.
\begin{lemma}
For nonnegative variables $x_1$, $x_2$, $y_1$ and $y_2$, we have:
\[ \left|\frac{x_1y_1}{1+x_1+y_1}-\frac{x_2y_2}{1+x_2+y_2}\right|\leq|x_1-x_2|+|y_1-y_2|.\]
\end{lemma}
\begin{proof}
We use the following inequality, which can be easily verified by
basic computations. For any $a\geq 0$, $b\geq 0$ and $c>0$ we have:
\begin{equation}
\left|\frac{a}{a+c}-\frac{b}{b+c}\right|\leq\frac{1}{c}|a-b|.
\end{equation}
Now the expression in Lemma 1 can be written as:\small{
\begin{IEEEeqnarray}{lll}
~\left|\frac{x_1y_1}{1{+}x_1{+}y_1}-\frac{x_2y_2}{1{+}x_2{+}y_2}\right|\nonumber\\
{\stackrel{\textmd{(a)}}{{\leq}}}\left|\frac{x_1y_1}{1{+}x_1{+}y_1}{-}\frac{x_1y_2}{1{+}x_1{+}y_2}\right|{+}
\left|\frac{x_1y_2}{1{+}x_1{+}y_2}{-}\frac{x_2y_2}{1{+}x_2{+}y_2}\right|\nonumber\\
=x_1\left|\frac{y_1}{y_1{+}{(}x_1{+}1{)}}{-}\frac{y_2}{y_2{+}{(}x_1{+}1{)}}\right|{+}
y_2\left|\frac{x_1}{x_1{+}{(}y_2{+}1{)}}{-}\frac{x_2}{x_2{+}{(}y_2{+}1{)}}\right|\nonumber\\
{\stackrel{(\textmd{b})}{{\leq}}}\frac{x_1}{x_1{+}1}\left|y_1-y_2\right|+\frac{y_2}{y_2{+}1}\left|x_1-x_2\right|
{\stackrel{(\textmd{c})}{{\leq}}}|x_1-x_2|+|y_1-y_2|,\nonumber
\end{IEEEeqnarray}}\normalsize
where (a) is the triangle inequality and (b) results from (II.1).
Finally (c) results from $\frac{x_1}{x_1{+}1}{<}1$ and
$\frac{y_2}{y_2{+}1}{<}1$, since $x_1$ and $y_2$ are nonnegative.
\end{proof}
\begin{lemma}
For the matrix $\mathbf{H}\in\mathds{C}^{p\times q}$ with
independent $\mathcal{CN}(0,1)$ entries, we have:
$\mathrm{E}\left\{\sum_{i}{\sigma_i^2}\right\}=pq$, where
$\sigma_i$'s are the singular values of $\mathbf{H}$.
\end{lemma}
\begin{proof}
Let $\mathbf{H}=[h_{ij}]$, where $h_{ij}\sim\mathcal{CN}(0,1)$. We
have:\begin{IEEEeqnarray}{lll}\mathrm{E}\left\{\sum_{i}{\sigma_i^2}\right\}&=&\mathrm{E}\left\{\textmd{Trace}(\mathbf{H}\mathbf{H}^H)\right\}\nonumber\\
&=&\mathrm{E}\left\{\sum_{i,j}{|h_{ij}|^2}\right\}=\sum_{i,j}{\mathrm{E}\left\{|h_{ij}|^2\right\}}=pq.\nonumber\end{IEEEeqnarray}\end{proof}
\begin{lemma}
Consider the codebook
$\mathbf{C}=\{\mathbf{w}_1,\mathbf{w}_2,\cdots,\mathbf{w}_N\}$ and
the matrix $\mathbf{H}$ with $\sigma_i$'s as its singular values.
For any unit vector $\mathbf{s}$ define
$\mathbf{s}_{_{\mathbf{C}}}\in\mathbf{C}$ as the closest vector in
codebook $\mathbf{C}$ to $\mathbf{s}$ and let
$d_{_{\mathbf{C}}}(\mathbf{s})\stackrel{def}{=}d(\mathbf{s},\mathbf{s}_{_{\mathbf{C}}})$,
where $d(\cdot,\cdot)$ is the distance function defined in (1).
Then, we have:
\[\left|\|\mathbf{H}\mathbf{s}\|^2-\|\mathbf{H}\mathbf{s}_{_{\mathbf{C}}}\|^2\right|\leq
2\left(\sum_{i}{\sigma_i^2}\right)d_{_{\mathbf{C}}}(\mathbf{s}),\]
\end{lemma}
\begin{proof}
For arbitrary unit vectors $\mathbf{u}$, $\mathbf{v}$ and
$\mathbf{w}$, we have the following from the triangle inequality:
\[|d(\mathbf{u},\mathbf{v})-d(\mathbf{v},\mathbf{w})|\leq d(\mathbf{u},\mathbf{w}).
\] On the other hand,
\[|d(\mathbf{u},\mathbf{v})+d(\mathbf{v},\mathbf{w})|\leq |d(\mathbf{u},\mathbf{v})|+|d(\mathbf{v},\mathbf{w})|\leq 2.
\] By multiplying the both sides of these inequalities we get:
\[\left|d^2(\mathbf{u},\mathbf{v})-d^2(\mathbf{v},\mathbf{w})\right|\leq
2d(\mathbf{u},\mathbf{w}).\] Considering the definition of the
distance function $d(\cdot,\cdot)$ in (1) we have:
\begin{equation}
\left|\left|\mathbf{u}^H\mathbf{v}\right|^2-\left|\mathbf{v}^H\mathbf{w}\right|^2\right|\leq
2d(\mathbf{u},\mathbf{w}).
\end{equation}

Now, if the right singular vectors of $\mathbf{H}$ are denoted by
$\mathbf{v}_i$'s, we have:
\begin{IEEEeqnarray}{lll}
\left|\|\mathbf{H}\mathbf{s}\|^2-\|\mathbf{H}\mathbf{w}\|^2\right|&=&\left|\sum_{i}{\sigma_i^2\left(\left|\mathbf{v}_i^H\mathbf{s}\right|^2-\left|\mathbf{v}_i^H\mathbf{w}\right|^2\right)}\right|\nonumber\\
&\leq&\sum_{i}{\sigma_i^2\left|\left|\mathbf{v}_i^H\mathbf{s}\right|^2-\left|\mathbf{v}_i^H\mathbf{w}\right|^2\right|},\nonumber
\end{IEEEeqnarray}
and by applying (II.2), we get:
\begin{equation}
\left|\|\mathbf{H}\mathbf{s}\|^2-\|\mathbf{H}\mathbf{w}\|^2\right|\leq
2\left(\sum_{i}{\sigma_i^2}\right)d(\mathbf{s},\mathbf{w}),
\end{equation}
The proof will be complete after substituting $\mathbf{w}$ in (II.3)
by $\mathbf{s}_{_{\mathbf{C}}}$.
\end{proof}

\begin{lemma}
Consider the codebook $\mathbf{C}(N,\delta)$ and the function
$d_{_{\mathbf{C}}}(\cdot)$ defined in Lemma 3. For the random vector
$\mathbf{s}\in\mathds{C}^m$ uniformly distributed on the unit sphere
we have:
\begin{equation}
\mathrm{E}\left\{d_{_{\mathbf{C}}}(\mathbf{s})\right\}\leq
1-N\left(\frac{\delta}{2}\right)^{2(m-1)}\left(1-\frac{\delta}{2}\right).\nonumber
\end{equation}
\end{lemma}
\begin{proof}
The proof is based on the arguments given in [5].
\end{proof}

\subsection{Proof of the Upper Bound in (22)}
The optimal unquantized SNR $\gamma^\star$ and the quantized scheme
SNR $\tilde{\gamma}$ are given in (17) and (21), which are repeated
here:
\begin{equation}
\gamma^\star=\frac{\gamma_1^\star\gamma_2^\star}{1+\gamma_1^\star+\gamma_2^\star},~~
\tilde{\gamma}=\frac{\tilde{\gamma_1}\tilde{\gamma_2}}{1+\tilde{\gamma_1}+\tilde{\gamma_2}},
\end{equation}
where $\gamma_1^\star$, $\gamma_2^\star$, $\tilde{\gamma}_1$ and
$\tilde{\gamma}_2$ are defined in (18) and (20). Clearly
$\gamma_1^\star>\tilde{\gamma}_1$ and
$\gamma_2^\star>\tilde{\gamma}_2$, and therefore,
$\gamma^\star>\tilde{\gamma}$. Our goal is to bound
$\gamma^\star-\tilde{\gamma}$. For this purpose, we need the
following definitions.
\begin{equation}
\gamma'_1{\stackrel{def}{{=}}}P_1{\|}\mathbf{H}_1\mathbf{b}_{_{\mathbf{C}_1}}{\|}^2,~~~
\gamma'_2{\stackrel{def}{{=}}}P_2{\|}\mathbf{H}_2\mathbf{g}_{_{\mathbf{C}_2}}{\|}^2,~~~
\gamma'{\stackrel{def}{{=}}}\frac{\gamma'_1\gamma'_2}{1{+}\gamma'_1{+}{\gamma'_2}},\nonumber\\
\end{equation}
where $\mathbf{b}_{_{\mathbf{C}_1}}$ is the closest vector in the
codebook $\mathbf{C}_1$ to $\mathbf{b}_1$, and
$\mathbf{g}_{_{\mathbf{C}_2}}$ is the closest vector in the codebook
$\mathbf{C}_2$ to $\mathbf{g}_1$. Note that, by the notation of
Section III, $\mathbf{b}_1$ and $\mathbf{g}_1$ are the strongest
right singular vectors of $\mathbf{H}_1$ and $\mathbf{H}_2$. By
considering the definitions of $\tilde{\gamma}_1$ and
$\tilde{\gamma}_2$ in (20) and the fact that
$\mathbf{b}_{_{\mathbf{C}_1}}\in\mathbf{C}_1$ and
$\mathbf{g}_{_{\mathbf{C}_2}}\in\mathbf{C}_2$, it is clear that
$\tilde{\gamma}_1>\gamma'_1$ and $\tilde{\gamma}_2>\gamma'_2$, and
therefore, $\tilde{\gamma}>\gamma'$. Hence, we can write:
\begin{IEEEeqnarray}{lll}
\gamma^\star-\tilde{\gamma}\leq\gamma^\star-\gamma'&=&\frac{\gamma_1^\star\gamma_2^\star}{1+\gamma_1^\star+\gamma_2^\star}-\frac{\gamma'_1\gamma'_2}{1+\gamma'_1+\gamma'_2}\nonumber\\
&\stackrel{\textmd{(a)}}{\leq}&\left(\gamma_2^\star-\gamma'_2\right)+\left(\gamma_1^\star-\gamma'_1\right),
\end{IEEEeqnarray}
where for (a) we have used Lemma 1. The terms on the right side of
(II.5) can be bounded as follows.

Noting the definitions of $\gamma_1^\star$, $\gamma'_1$ we have:
\[\gamma_1^\star{-}\gamma'_1{=}P_1\left(\|\mathbf{H}_1\mathbf{\mathbf{b}_1}\|^2{-}\|\mathbf{H}_1\mathbf{b}_{_{\mathbf{C}_1}}\|^2\right)
\stackrel{\textmd{(b)}}{\leq}2P_1\left(\sum_i{\phi_i^2}\right){d_{_{\mathbf{C}_1}}{(}\mathbf{b_1}{)}},\]
where for (b) we have used Lemma 3, and $\phi_i$'s are singular
values of $\mathbf{H}_1$. The term $\gamma_2^\star{-}\gamma'_2$ can
be similarly bounded. Combining these bounds with (II.5), we get the
following upper bound:
\begin{equation}\gamma^\star-\tilde{\gamma}\leq2\left(\sum_{i}{\phi_i^2}\right)d_{_{\mathbf{C}_1}}(\mathbf{b_1})+
2\left(\sum_{i}{\psi_i^2}\right)d_{_{\mathbf{C}_2}}(\mathbf{g_1}),\end{equation}
where $\psi_i$'s are singular values of $\mathbf{H}_2$. Noting that
the singular vectors $\mathbf{b}_1$ and $\mathbf{g}_1$ are uniformly
distributed on the unit spheres (of the corresponding dimension) and
are independent of the singular values, we can apply Lemma 2 and 4
to (II.6) to achieve the upper bound in (22).

\subsection{Proof of the Upper Bound in (30)}
Define:
\[\gamma(\mathbf{s}_1,\mathbf{s}_2)\stackrel{def}{=}\frac{\gamma_1(\mathbf{s}_1)\gamma_2(\mathbf{s}_2)}{1+\gamma_1(\mathbf{s}_1)+\gamma_2(\mathbf{s}_2)}+\gamma_0(\mathbf{s}_1),\]
where
$\gamma_i(\mathbf{s})\stackrel{def}{=}P_i\|\mathbf{H}_i\mathbf{s}\|^2$,
for $i=0,1,2$. With these definitions, the SNR of the optimal
unquantized scheme $\gamma^\star$ and the SNR of the quantized
scheme $\tilde{\gamma}$ can be expressed as:
\begin{IEEEeqnarray}{lll}
\gamma^\star&=&\max_{\substack{\|\mathbf{s}_1\|=1\\ \|\mathbf{s}_2\|=1}}{\gamma(\mathbf{s}_1,\mathbf{s}_2)}={\gamma(\mathbf{s}^\star,\mathbf{g}_1)}=\frac{\gamma_1^\star\gamma_2^\star}{1+\gamma_1^\star+\gamma_2^\star}+\gamma_0^\star\nonumber\\
\tilde{\gamma}&=&\max_{\substack{\mathbf{w}_1\in\mathbf{C}_1\\\mathbf{w}_2\in\mathbf{C}_2}}{\gamma(\mathbf{w}_1,\mathbf{w}_2)}={\gamma(\tilde{\mathbf{s}},\tilde{\mathbf{g}})},
\end{IEEEeqnarray}
where $\mathbf{g}_1$ is the strongest right singular vector of
$\mathbf{H}_2$, and $\mathbf{s}^\star$, $\tilde{\mathbf{s}}$ and
$\tilde{\mathbf{g}}$ are defined in (25), (29) and (28). Also
$\gamma_0^\star=\gamma_0(\mathbf{s}^\star)$,
$\gamma_1^\star=\gamma_1(\mathbf{s}^\star)$, and
$\gamma_2^\star=\gamma_2(\mathbf{g}_1)$.

Our goal is to bound the SNR loss $\gamma^\star-\tilde{\gamma}$. For
this purpose, we need the following definitions.
\begin{IEEEeqnarray}{lll}
\gamma'\stackrel{def}{=}\frac{\gamma'_1\gamma'_2}{1+\gamma'_1+\gamma'_2}+\gamma'_0,\nonumber\\
\gamma'_0\stackrel{def}{=}\gamma_0(\mathbf{s}^\star_{_{\mathbf{C}_1}}),~~
\gamma'_1\stackrel{def}{=}\gamma_1(\mathbf{s}^\star_{_{\mathbf{C}_1}}),~~
\gamma'_2\stackrel{def}{=}\gamma_2(\mathbf{g}_{_{\mathbf{C}_2}})\nonumber,
\end{IEEEeqnarray}
where $\mathbf{s}^\star_{_{\mathbf{C}_1}}\in\mathbf{C}_1$ is the
closest vector in the codebook $\mathbf{C}_1$ to $\mathbf{s}^\star$,
and $\mathbf{g}_{_{\mathbf{C}_2}}\in\mathbf{C}_2$ is the closest
vector in the codebook $\mathbf{C}_2$ to $\mathbf{g}_1$.

Noting the above definitions, it is clear that
$\gamma^\star\geq\tilde{\gamma}\geq\gamma'$ and we can write:
\begin{IEEEeqnarray}{lll}
\gamma^\star{-}\tilde{\gamma}\leq\gamma^\star{-}\gamma'
\leq\left|\frac{\gamma_1^\star\gamma_2^\star}{1{+}\gamma_1^\star{+}\gamma_2^\star}{-}\frac{\gamma'_1\gamma'_2}{1{+}\gamma'_1{+}\gamma'_2}\right|+\left|\gamma_0^\star{-}\gamma'_0\right|\nonumber\\
~\stackrel{\textmd{(a)}}{\leq}\left|\gamma_2^\star-\gamma'_2\right|+\left|\gamma_1^\star-\gamma'_1\right|+\left|\gamma_0^\star-\gamma'_0\right|,\nonumber\\
~{\stackrel{\textmd{(b)}}{=}}P_2\left|\|\mathbf{H}_2\mathbf{\mathbf{g}_1}\|^2-\|\mathbf{H}_2\mathbf{g}_{_{\mathbf{C}_2}}\|^2\right|
+P_1\left|\|\mathbf{H}_1\mathbf{\mathbf{s}^\star}\|^2-\|\mathbf{H}_1\mathbf{s}^\star_{_{\mathbf{C}_1}}\|^2\right|\nonumber\\
~+P_0\left|\|\mathbf{H}_0\mathbf{\mathbf{s}^\star}\|^2-\|\mathbf{H}_0\mathbf{s}^\star_{_{\mathbf{C}_1}}\|^2\right|\nonumber\\
~\stackrel{\textmd{(c)}}{\leq}~2P_2\left(\sum_{i}{\psi_i^2}\right)d_{_{\mathbf{C}_2}}(\mathbf{g}_1)+2P_1\left(\sum_{i}{\phi_i^2}\right)d_{_{\mathbf{C}_1}}(\mathbf{s}^\star)\nonumber\\
~+2P_0\left(\sum_{i}{\nu_i^2}\right)d_{_{\mathbf{C}_1}}(\mathbf{s}^\star),
\end{IEEEeqnarray}
where we have used Lemma 1 for (a). In (b),
$\{\gamma_i^\star\}_{i=0}^2$ and $\{\gamma'_i\}_{i=0}^2$ have been
replaced by their definitions. Finally, (c) results from Lemma 3.

We know from Appendix I, that $\mathbf{s}^\star$ is uniformly
distributed on the unit sphere and is independent of the eigenvalues
$\nu_i$'s and $\phi_i$'s. The same argument holds for the singular
vector $\mathbf{g}_1$ and the singular values $\psi_i$'s.
Considering this, we can take expectation from both sides of (II.8)
and use Lemma 1 and Lemma 4 to achieve the upper bound in (30).

\subsection{Proof of the Upper Bound in (32)}
As in Appendix II.B, the SNR of the optimal unquantized is given by:
\[\gamma^\star=\max_{\substack{{\|\mathbf{s}_1\|=1}\\{\|\mathbf{s}_2\|=1}}}\gamma(\mathbf{s}_1,\mathbf{s}_2)=\gamma(\mathbf{s}^\star,\mathbf{g}_1),\]
where $\mathbf{s}^\star$, $\mathbf{g}_1$ and the function
$\gamma(\cdot,\cdot)$ are defined in Appendix II.B. As described in
Section IV.B.2, the quantized beamforming vectors are determined
from:
\begin{equation}
\tilde{\mathbf{g}}=\arg\max_{\mathbf{w}\in{\mathbf{C}_2}}{\gamma_1(\mathbf{w})},~~~
\tilde{\mathbf{s}}^\star=\arg\max_{\mathbf{w}\in{\mathbf{C}_1}}\chi(\mathbf{w},\tilde{\mathbf{g}}),
\end{equation}
where
\begin{equation}
\chi(\mathbf{s}_1,\mathbf{s}_2)\stackrel{def}{=}\frac{\gamma_1(\mathbf{s}_1)\gamma_2(\mathbf{s}_2)}{1{+}\gamma_1(\mathbf{s}_1){+}\gamma_2(\mathbf{s}_2)}{+}P_0
\sum_{i}{\nu_i^2\left|\tilde{\mathbf{e}}_i^H\mathbf{s}_1\right|^2}.
\end{equation}
In (II.10), $\nu_i$'s are the singular values of $\mathbf{H}_0$ and
$\tilde{\mathbf{e}}_i$'s are the quantized version of
$\mathbf{e}_i$'s which are the right singular vectors of
$\mathbf{H}_0$. The SNR value resulted from the choices in (II.9)
is:
\begin{equation}
\tilde{\gamma}^\star=\gamma(\tilde{\mathbf{s}}^\star,\tilde{\mathbf{g}}).
\end{equation}

Our goal is to bound $\gamma^\star-\tilde{\gamma}^\star$. For this
purpose, we need the following definitions from Appendix II.B:
\begin{IEEEeqnarray}{lll}
\tilde{\mathbf{s}}=\arg\max_{\mathbf{w}\in{\mathbf{C}_1}}\gamma(\mathbf{w},\tilde{\mathbf{g}})\nonumber\\
\tilde{\gamma}=\gamma(\tilde{\mathbf{s}},\tilde{\mathbf{g}}).
\end{IEEEeqnarray}

The SNR loss $\gamma^\star-\tilde{\gamma}^\star$ can be expressed
as:
\begin{equation}
\gamma^\star-\tilde{\gamma}^\star=\left(\gamma^\star-\tilde{\gamma}\right)+\left(\tilde{\gamma}-\tilde{\gamma}^\star\right).
\end{equation}
The first term has already been bounded in Appendix II.B. To bound
the second term we will need the result proven in Lemma 5 (at the
end of this section). Let
$\theta=2P_0\sum_{i}{\nu_i^2d_{_{\mathbf{C_0}}}(\mathbf{e}_i)}$,
then we have:
\begin{IEEEeqnarray}{lll}
\tilde{\gamma}=\gamma(\tilde{\mathbf{s}},\tilde{\mathbf{g}})\stackrel{\textmd{(a)}}{\leq}
\chi(\tilde{\mathbf{s}},\tilde{\mathbf{g}})+\theta \stackrel{\textmd{(b)}}{\leq}\chi(\tilde{\mathbf{s}}^\star,\tilde{\mathbf{g}})+\theta\nonumber\\
~~~~~~~~~~~~~~\stackrel{\textmd{(c)}}{\leq}\gamma(\tilde{\mathbf{s}}^\star,\tilde{\mathbf{g}})+2\theta=\tilde{\gamma}^\star+2\theta,
\end{IEEEeqnarray}
where in (a) and (c) we have used Lemma 5, and (b) results from
(II.9) and the fact that $\tilde{\mathbf{s}}\in\mathbf{C}_1$. By
combining (II.14), (II.13) and (II.8) we get the following upper
bound:
\begin{IEEEeqnarray}{lll}
\gamma^\star-\tilde{\gamma}^\star&\leq&2P_2\left(\sum_{i}{\psi_i^2}\right)d_{_{\mathbf{C}_2}}(\mathbf{g}_1)\nonumber\\
&+&2\left(P_1\left(\sum_{i}{\phi_i^2}\right)+P_0\left(\sum_{i}{\nu_i^2}\right)\right)d_{_{\mathbf{C}_1}}(\mathbf{s}^\star)\nonumber\\
&+&4P_0\sum_{i}{\nu_i^2d_{_{\mathbf{C}_0}}(\mathbf{e}_i)}.
\end{IEEEeqnarray}
From Appendix I, $\mathbf{s}^\star$ is uniformly distributed on the
unite sphere and is independent of the singular values $\phi_i$'s
and $\nu_i$'s. The same argument holds for the singular vectors
$\mathbf{g}_1$ and $\mathbf{e}_i$'s and the corresponding singular
values $\psi_i$'s and $\nu_i$'s. By considering these facts and
taking the expectation of both sides of (II.15) and using Lemma 1
and 4, we get the upper bound in (32).

\begin{lemma}
For any unit vector $\mathbf{s}$, we have:
\[\left|\gamma(\mathbf{s},\tilde{\mathbf{g}})-\chi(\mathbf{s},\tilde{\mathbf{g}})\right|\leq
2P_0\sum_{i}{\nu_i^2d_{_{\mathbf{C_0}}}(\mathbf{e}_i)}.\]
\end{lemma}
\begin{proof}
Noting the definition of $\gamma(\cdot,\cdot)$ in Appendix II.B,
\[
\gamma(\mathbf{s},\tilde{\mathbf{g}})=\frac{\gamma_1(\mathbf{s})\gamma_2(\tilde{\mathbf{g}})}{1+\gamma_1(\mathbf{s})+\gamma_2(\tilde{\mathbf{g}})}+
P_0\sum_{i}{\nu_i^2\left|\mathbf{e}_i^H\mathbf{s}\right|^2}.
\]
Therefore,
\begin{IEEEeqnarray}{lll}
\left|\gamma(\mathbf{s},\tilde{\mathbf{g}})-\chi(\mathbf{s},\tilde{\mathbf{g}})
\right|&=&P_0\left|\sum_{i}{\nu_i^2\left(\left|\mathbf{e}_i^H\mathbf{s}\right|^2-\left|\tilde{\mathbf{e}}_i^H\mathbf{s}\right|^2\right)}\right|\nonumber\\
&\stackrel{\textmd(a)}{\leq}&2P_0\sum_{i}{\nu_i^2}d(\mathbf{e}_i,\tilde{\mathbf{e}}_i),\nonumber
\end{IEEEeqnarray}
where in (a), we have used (II.1) in Lemma 3. Noting that
$\tilde{\mathbf{e}}_i$'s are by definition the closest vectors in
$\mathbf{C}_0$ to $\tilde{\mathbf{e}}_i$'s, we have
$d(\mathbf{e}_i,\tilde{\mathbf{e}}_i)=d_{_{\mathbf{C}_0}}(\mathbf{e}_i)$
and the proof is complete.
\end{proof}

\section{Comparison of the Optimal and Modified Unquantized Schemes}
In this appendix the following lemma will be used to bound the SNR
loss of the modified unquantized scheme with respect to the optimal
unquantized scheme.
\begin{lemma}
Consider the SVD $\mathbf{H}=\mathbf{U}\mathbf{\Sigma}\mathbf{V}^H$
for an arbitrary matrix $\mathbf{H}\in\mathds{C}^{l\times n}$, where
$\mathbf{U}\in\mathcal{U}^l$,
$\mathbf{V}=[\mathbf{v}_1|\cdots|\mathbf{v}_n]\in\mathcal{U}^n$, and
$\mathbf{\Sigma}=\texttt{diag}_{l\times
n}(\sigma_1,\sigma_2,\cdots,\sigma_r)$, where $r=\min\{l,n\}$. Then
for any unit vector $\mathbf{s}$ we have:
\[\sigma_1^2\left|\mathbf{v}_1^H\mathbf{s}\right|^2\leq
\|\mathbf{H}\mathbf{s}\|^2 \leq
\sigma_1^2\left|\mathbf{v}_1^H\mathbf{s}\right|^2+\sigma_2^2.
\]
\end{lemma}
\begin{proof}
Note that
$\|\mathbf{H}\mathbf{s}\|^2=\sum_{i=1}^{n}{\sigma_i^2\left|\mathbf{v}_i^H\mathbf{s}\right|^2}$.
The left side inequality in Lemma 6 is obvious, since
$\sigma_i^2\left|\mathbf{v}_i^H\mathbf{s}\right|^2\geq 0$ for $i>1$.
The right side inequality can be proven as follows:
\begin{IEEEeqnarray}{lll}
\|\mathbf{H}\mathbf{s}\|^2&=&\sigma_1^2\left|\mathbf{v}_1^H\mathbf{s}\right|^2+\sum_{i>1}{\sigma_i^2\left|\mathbf{v}_i^H\mathbf{s}\right|^2}\nonumber\\
&\stackrel{\textmd{(a)}}{\leq}&\sigma_1^2\left|\mathbf{v}_1^H\mathbf{s}\right|^2+\sigma_2^2\sum_{i>1}{\left|\mathbf{v}_i^H\mathbf{s}\right|^2}\nonumber\\
&\stackrel{\textmd{(b)}}{\leq}&\sigma_1^2\left|\mathbf{v}_1^H\mathbf{s}\right|^2+\sigma_2^2\sum_{i=1}^{n}{\left|\mathbf{v}_i^H\mathbf{s}\right|^2}
\stackrel{\textmd{(c)}}{\leq}\sigma_1^2\left|\mathbf{v}_1^H\mathbf{s}\right|^2+\sigma_2^2,\nonumber
\end{IEEEeqnarray}
where (a) results from $\sigma_2\geq\sigma_i$ for $i>1$. In (b) we
are adding the nonnegative term
$\sigma_2^2\left|\mathbf{v}_1^H\mathbf{s}\right|^2$ and (c) results
from
\[\sum_{i=1}^{n}{\left|\mathbf{v}_i^H\mathbf{s}\right|^2}=\|\mathbf{V}^H\mathbf{s}\|^2=\mathbf{s}^H\mathbf{V}\mathbf{V}^H\mathbf{s}=\mathbf{s}^H\mathbf{s}=\|\mathbf{s}\|^2=1,\]
since $\mathbf{V}$ is a (square) unitary matrix.
\end{proof}

Considering the definition of the function $\gamma(\cdot,\cdot)$ in
Appendix II.B, the SNR of the optimal unquantized is given by:
\[\gamma^\star=\max_{\substack{{\|\mathbf{s}_1\|=1}\\{\|\mathbf{s}_2\|=1}}}\gamma(\mathbf{s}_1,\mathbf{s}_2)=\gamma(\mathbf{s}^\star,\mathbf{g}_1),\]
where $\mathbf{g}_1$ is the strongest right singular vector of
$\mathbf{H}_2$ and
$\mathbf{s}^\star=\arg\max_{\|\mathbf{s}\|=1}\gamma(\mathbf{s},\mathbf{g}_1)$.

On the other hand the Tx beamforming vector of the modified
unquantized scheme is determined by:
\begin{equation}
\mathbf{s}_{_{modified}}=\arg\max_{\|\mathbf{s}\|=1}{\xi(\mathbf{s},\mathbf{g}_1)},
\end{equation}
where
\[\xi(\mathbf{s}_1,\mathbf{s}_2)\stackrel{def}{=}\frac{\gamma_1(\mathbf{s}_1)\gamma_2({\mathbf{s}_2})}{1+\gamma_1(\mathbf{s}_1)+\gamma_2({\mathbf{s}_2})}+
P_0{\nu_1^2\left|\mathbf{e}_1^H\mathbf{s}_1\right|^2}.
\]
Here $\nu_1$ and $\mathbf{e}_1$ are the largest singular value and
strongest right singular vector of $\mathbf{H}_0$, respectively. The
corresponding SNR of the modified scheme is:
\[\gamma_{_{modified}}=\gamma(\mathbf{s}_{_{modified}},\mathbf{g}_1).\]

Noting the definitions of $\gamma(\cdot,\cdot)$ and
$\xi(\cdot,\cdot)$ and using Lemma 6, we have the following for any
unit vector $\mathbf{s}$:
\begin{equation}\gamma(\mathbf{s},\mathbf{g}_1)\leq\xi(\mathbf{s},\mathbf{g}_1)+P_0\nu_2^2,\end{equation}
where $\nu_2$ is the second largest singular value of
$\mathbf{H}_0$. Taking the maximum of the both sides of (III.2) over
the unit sphere, we get:
\begin{IEEEeqnarray}{lll}
\gamma^\star&=&\gamma(\mathbf{s}^\star,\mathbf{g}_1)\leq
\xi(\mathbf{s}_{_{modified}},\mathbf{g}_1)+P_0\nu_2^2\nonumber\\
&\stackrel{\textmd{(a)}}{\leq}&\gamma(\mathbf{s}_{_{modified}},\mathbf{g}_1)+P_0\nu_2^2=\gamma_{_{modified}}+P_0\nu_2^2,
\end{IEEEeqnarray}
where (a) results from the fact that
$\xi(\mathbf{s}_1,\mathbf{s}_2)$ is globally upper bounded by
$\gamma(\mathbf{s}_1,\mathbf{s}_2)$ for any $\mathbf{s}_1$ and
$\mathbf{s}_2$ (Note the first inequality in Lemma 6 and the
definitions of $\gamma(\cdot,\cdot)$ and $\xi(\cdot,\cdot)$). Taking
expectation of both sides of (III.3), we get:
\begin{equation}
\mathrm{E}\{\gamma^\star\}-\mathrm{E}\{\gamma_{_{modified}}\}\leq
P_0\mathrm{E}\{\nu_2^2\}.
\end{equation}

On the other hand,
\[
\gamma^\star=\max_{\substack{{\|\mathbf{s}_1\|=1}\\{\|\mathbf{s}_2\|=1}}}\gamma(\mathbf{s}_1,\mathbf{s}_2)\geq\gamma(\mathbf{e}_1,\mathbf{g}_1)\geq
P_0\nu_1^2,
\]and therefore, $\mathrm{E}\{\gamma^\star\}\geq
P_0\mathrm{E}\{\nu_1^2\}$. Combining this with (III.4), we get the
following upper bound.
\[
\frac{\mathrm{E}\{\gamma^\star\}}{\mathrm{E}\{\gamma_{_{modified}}\}}\leq
1+\frac{\mathrm{E}\{\nu_2^2\}}{\mathrm{E}\{\nu_1^2\}},
\]
or
\[\mathrm{E}\{\gamma^\star\}_{\textmd{dB}}-\mathrm{E}\{\gamma_{_{modified}}\}_{\textmd{dB}}\leq
10\log_{10}\left(1+\frac{\mathrm{E}\{\nu_2^2\}}{\mathrm{E}\{\nu_1^2\}}\right).\]
For Rayleigh channel matrix $\mathbf{H}_0\in\mathds{C}^{3\times 3}$,
this upper bound is equal to $1.24$dB.

\end{document}